\documentclass{article}

\pdfoutput=1
\usepackage{amsmath}
\usepackage{amsfonts}
\usepackage{amssymb}
\usepackage{graphicx}
\usepackage[round]{natbib} 
\usepackage[lined,boxed, norelsize]{algorithm2e}
\usepackage{animate}
\usepackage{amsthm}
\usepackage{multirow}
\usepackage{hyperref}
\usepackage{lscape}
\usepackage{rotating}
\usepackage{animate}
\usepackage{setspace}

\begin{document}

\title{Estimation and approximation in nonlinear dynamic systems using quasilinearization}
\author{Gianluca Frasso, Jonathan Jaeger \& Philippe Lambert}

\maketitle


\begin{abstract}
Nonlinear (systems of) ordinary differential equations (ODEs) are common tools in the analysis of complex one-dimensional dynamic systems. In this paper we propose a smoothing approach regularized by a quasilinearized ODE-based penalty in order to approximate the state functions and estimate the parameters defining nonlinear differential systems from noisy data. Within the quasilinearized spline based framework, the estimation process reduces to a conditionally linear problem for the optimization of the spline coefficients. Furthermore, standard ODE compliance parameter(s) selection criteria are easily applicable and conditions on the state function(s) can be eventually imposed using soft or hard constraints. The approach is illustrated on real and simulated data. 
\\ 
\\
\textbf{Keywords:} Differential conditions; Nonlinear differential equations; Penalized splines; Quasilinearization
\end{abstract}

\section{Introduction}
\label{section:introduction}
The analysis of one dimensional dynamic systems represents a crucial task in many scientific fields. This kind of problem is usually tackled invoking differential calculus tools by defining (a system of) ordinary differential equation(s) which (analytic or numerical) solution(s) appropriately describes the observed dynamics.\\
In many circumstances, the parameters involved in the differential model synthesizing the observed phenomenon are not known. In these cases statistical procedures are applied to estimate them from noisy data. In general it is assumed that changes in the states $\boldsymbol{x}\left(t\right) \in \mathbb{R}^{d}$ of a dynamic system are governed by a set of differential equations:
\begin{equation}
\displaystyle\frac{d\boldsymbol{x}}{dt}\left(t\right) = \boldsymbol{f}\left(t, \boldsymbol{x}, \boldsymbol{\theta}\right), t\in\left[0;T\right],
\label{eq:def_eq_diff}
\end{equation}
where $\boldsymbol{f}$ is a known nonlinear function in the states $\boldsymbol{x}$ and $\boldsymbol{\theta} \in \mathbb{R}^{q}$ an unknown vector of parameters. It is assumed that only a subset $\mathcal{J}\subset\left\{1,\ldots,d\right\}$ of the $d$ state functions $\boldsymbol{x}$ are observed at time point $t_{jk}$, $j \in \mathcal{J}$, $k=1,\ldots,n_{j}$ with additive measurement error $\epsilon_{jk}$. We denote by $y_{jk}=x_{j}\left(t_{jk}\right) + \tau_{j}^{-1 / 2}\epsilon_{jk}$ the corresponding measurement. For simplicity, we assume that the error terms $\epsilon_{jk}$ are distributed according to a standardized Gaussian distribution.
The goal is to estimate the ODE parameters $\boldsymbol{\theta}$, the precision of measurements $\boldsymbol{\tau} = \mbox{vec}\left(\tau_{j}, j \in \mathcal{J}\right)$ and to approximate the solution of the ODE model given in Eq.~\eqref{eq:def_eq_diff} from the (noisy) observations.\\
Many approaches have been proposed in the literature to estimate these sets of unknowns. The first reference in this field is probably \cite{hotelling1927} that tackled this task from a regression point of view. Following this idea, \cite{biegleretal1986} proposed a nonlinear least squares approach for the estimation of the ODE parameters using suitable numerical schemes to approximate the state function(s). This approach suffers of some drawbacks. It can be computationally demanding to deal with complex differential systems and the quality of the final estimates can be severely influenced by possible error propagation effects (related to the adopted numerical scheme). Finally these techniques become unpractical for inferential purposes in complex dynamics.\\
For these reasons the penalized smoothing collocation approach of \cite{ramsayetal2007} represents an attractive alternative. This estimation procedure consists in a compromise between a flexible (regression based) description of the recorded measurements and the (numerical) solution of the hypothesized differential system obtained through a collocation scheme. In particular, this framework exploits a B-spline approximation of the state function. The flexibility induced by the B-spline approximation is then counterbalanced by a penalty related to the differential model. The ODE-penalty term is defined as the integral of the ODE operator evaluated at the B-spline approximation of the state function. It measures the fidelity of the B-spline approximation to the ODE model. An ODE-compliance parameter balances between the fidelity to the data and the compliance of the B-spline approximation to the ODE model. This approach can be adopted both in case of linear and nonlinear differential equations.\\
In contrast to the linear cases, estimation in nonlinear systems using the approach by \cite{ramsayetal2007} requires a nonlinear least squares step and the application of the implicit function theorem both for point and interval estimates of the ODE parameters. Furthermore, even if the compliance parameter still tunes the balance between data fitting and ODE model fidelity, the application of standard selection procedure can be unpractical to deal with nonlinear ODE systems.\\
In this paper, we overcome these challenges by introducing a quasilinearization scheme \citep{bellmanandkalaba1965} in the estimation procedure. Our quasilinearized ODE-P-spline (QL-ODE-P-spline) approach can be viewed as a convenient generalization of the P-spline collocation procedure of \cite{ramsayetal2007} as it permits an explicit link between the spline coefficients and the unknown differential parameters. Furthermore, standard approaches for selecting/optimizing the ODE-compliance parameters become directly applicable.\\
In this paper we consider two differential problems as working examples to illustrate our proposal and evaluate its performances and robustness through simulations: 1) a simple first order ODE with initial value condition having a closed form solution; 2) the well known Van der Pol system of ODEs. The \textit{first order initial value problem}
\begin{equation}
\left\{
\begin{array}{r c l}
\displaystyle\frac{dx}{dt}\left(t\right) & = & \left(\theta_{1} - 2 \theta_{2} t\right) x ^ {2}\left(t\right)\\
x\left(0\right) & = & x_{0},
\end{array}
\right.
\label{eq:first_order_example}
\end{equation}
with explicit solution 
\begin{equation}
x\left(t\right) = \displaystyle\frac{1}{\theta_{2}t^{2} - \theta_{1}t + x_{0}^{-1}},
\label{eq:solution_first_order}
\end{equation}
appears particularly attractive as it has a closed form solution, enabling an objective comparison of the merits of our approach with the nonlinear least squares one. The second working example is represented by the \textit{Van der Pol system}:
\begin{equation}
\left\{
\begin{array}{r c l}
\displaystyle\frac{d x_{1}}{dt}\left(t\right) & = & \theta \left(x_{1}\left(t\right) - \displaystyle\frac{1}{3}x_{1}^{3}\left(t\right) - x_{2}\left(t\right)\right)\\
\displaystyle\frac{d x_{2}}{dt}\left(t\right) & = & \displaystyle\frac{1}{\theta} x_{1}\left(t\right).
\end{array}
\right.
\label{eq:second_order_example}
\end{equation}
In this system, the state function $x_{1}\left(t\right)$ describes a non-conservative oscillator with nonlinear damping. In particular, it has been proposed to describe the dynamics of current charge in a nonlinear electronic oscillator circuit \citep{vanderpol1926}. No exact solution can be derived for non-zero $\theta$ parameter. This model also found applications in other fields such as biology and seismology. \\
This paper is organized as follows. In Section~\ref{section:generalized_profiling_estimation_in_a_nutshell} the key features of the generalized profiling estimation procedure proposed by \cite{ramsayetal2007} are briefly recalled. In Section~\ref{section:quasilinearized_ode_p_spline_approach}, the introduction of a quasilinearization step in the profiling framework is illustrated to highlight the strengths of our QL-ODE-P-spline method. 
Its properties are studied using simulations in Section~\ref{section:simulation}. We conclude the paper with the analysis of two challenging datasets in Section~\ref{section:application} followed by a discussion in Section~\ref{section:discussion}.
\section{Generalized profiling estimation in a nutshell}
\label{section:generalized_profiling_estimation_in_a_nutshell}
In this section, the key steps of the profiling procedure proposed by \cite{ramsayetal2007} are briefly reminded. First of all, each state function involved in the system of (nonlinear) differential equations is approximated using a B-spline basis function expansion:
\begin{eqnarray*}
\widetilde{x_{j}}\left(t\right) &=& \sum_{k = 1}^{K_{j}} b_{jk}\left(t\right) \alpha_{jk},\\
& = & \left[\boldsymbol{b}_{j} \left( t \right) \right]^{\top} \boldsymbol{\alpha}_{j}
\label{eq:b_spline_approx2}
\end{eqnarray*}
where $\boldsymbol{\alpha}_{j}$ is a $K_{j}$-vector of spline coefficients and $\boldsymbol{b}_{j}\left(t\right)$ is the $K_{j}$-vector of B-spline basis functions evaluated at time $t$. The variation of the B-spline approximations are shrunk by introducing an ODE-based penalty. More precisely, for each differential equation of the system, a penalty term is introduced:
\begin{equation}
{PEN}_{j}\left(\boldsymbol{\alpha} | \boldsymbol{\theta} \right) = \int{\left(\frac{d\widetilde{x}_{j}}{dt}\left(t\right) - f_{j}\left(t, \widetilde{\boldsymbol{x}}, \boldsymbol{\theta}\right)\right)^{2}dt},
\label{eq:pen_j}
\end{equation}
where $\boldsymbol{\alpha} = \mbox{vec}\left(\boldsymbol{\alpha}_{j}; j=1,\ldots,d\right)$. The full fidelity measure to the ODE-model is then defined as:
\begin{equation}
{PEN}\left(\boldsymbol{\alpha} | \boldsymbol{\theta}, \boldsymbol{\gamma}\right) = \sum_{j = 1}^{d}{\gamma_{j}{PEN}_{j}\left(\boldsymbol{\alpha}\right)}.
\label{eq:global_pen}
\end{equation}
The compromise between data fitting and model fidelity is tuned by the ODE-compliance parameters $\boldsymbol{\gamma}$. For values of $\gamma_{j}$ close to zero, the final estimates
tend to overfit the data by satisfying nearly exclusively the goodness of fit criterion. For $\gamma_{j} \rightarrow \infty$, the final approximation tends to mimic the solution of the ODE model defining the penalty.\\
The ODE parameters and the spline coefficients are estimated by profiling the likelihood. This approach requires to optimize the choice of the ODE parameters by considering the spline coefficients as nuisance parameters. For given values of the ODE parameters $ {\tau}_{j \in \mathcal{J}}$, of the precisions of measurement $\left\{ \tau \right\}_{j=1}^{J}$ and of the ODE-compliance parameters $\boldsymbol{\gamma}$, the optimal spline coefficients are found as the maximizer of:
\begin{equation}\label{eq:goodness_of_fit_profiling}
J\left(\boldsymbol{\alpha}|\boldsymbol{\theta}, \boldsymbol{\gamma}, \boldsymbol{\tau}, \boldsymbol{y}\right) = \sum_{j\in \mathcal{J}}{\left\{\frac{n_{j}}{2}\log\left(\tau_{j}\right) - \frac{\tau_{j}}{2}\left\|\boldsymbol{y}_{j} - \widetilde{x}_{j}\left(\boldsymbol{t}\right)\right\|^{2}\right\}} - \frac{1}{2}{PEN}\left(\boldsymbol{\alpha}|\boldsymbol{\theta}, \boldsymbol{\gamma}\right).
\end{equation}
In a second step, given the current value of the ODE compliance parameters, the vector $\boldsymbol{\theta}$ is estimated together with the precisions of measurement optimizing the following data fitting criterion:
\begin{equation}
H\left(\boldsymbol{\theta}, \boldsymbol{\tau}|\boldsymbol{y}, \widehat{\boldsymbol{\alpha}} \right) = \sum_{j \in \mathcal{J}}{\left\{\frac{n_{j}}{2}\log\left(\tau_{j}\right) - \frac{\tau_{j}}{2}\left\|\boldsymbol{y}_{j} - \boldsymbol{B}_{j}\widehat{\boldsymbol{\alpha}}_{j}\left(\boldsymbol{\theta}, \boldsymbol{\tau}, \boldsymbol{\gamma}, \boldsymbol{y}  \right)\right\|^{2}\right\}},
\label{eq:ODE_crit}
\end{equation}
with $\boldsymbol{B}_{j}$ the B-spline matrix defined for the $j$th state function.\\
As the definition of the penalty term ${PEN}\left(\boldsymbol{\alpha} | \boldsymbol{\theta} , \boldsymbol{\gamma}\right)$ is based on (a system of) nonlinear differential equation(s), the 
minimizer of $J\left(\boldsymbol{\alpha}|\boldsymbol{\theta}, \boldsymbol{\gamma}, \boldsymbol{\tau}, \boldsymbol{y}\right)$ cannot be found analytically. This makes the dependence of $H$ on $\boldsymbol{\theta}$ and $\boldsymbol{\tau}$ partially implicit. Therefore, the function $H$ can only be optimized combining a non linear least squares solver with the implicit function theorem. Similar arguments are involved in the approximation of the variance-covariance matrix of $\widehat{\boldsymbol{\theta}}$ (see \cite{ramsayetal2007} for details). The selection of the optimal ODE-compliance parameters is a non trivial task as common smoothing parameter selection criteria cannot be applied.
\section{Quasilinearized ODE-P-spline approach}
\label{section:quasilinearized_ode_p_spline_approach}
When nonlinearity becomes an issue for the estimation process, it is convenient to approximate nonlinear functions with linear surrogates (obtained for example through first order Taylor expansion). Quasilinearization \citep{bellmanandkalaba1965} generalizes this idea to approximate the solution of differential equations. In the generalized profiling estimation procedure, provided that differential equations are linear, the estimation of $\boldsymbol{\theta}$ and $\tau$ can be made by iterating simple least squares procedures with ODE compliance parameters selected using standard procedures. From the discussion at the end of Section~\ref{section:generalized_profiling_estimation_in_a_nutshell} it appears that dealing with nonlinear ODEs is more challenging. Our main suggestion is to amend the estimation procedure using quasilinearization.
\subsection{Quasilinearization of the penalty}
\begin{figure}
\includegraphics[width = 1\textwidth]{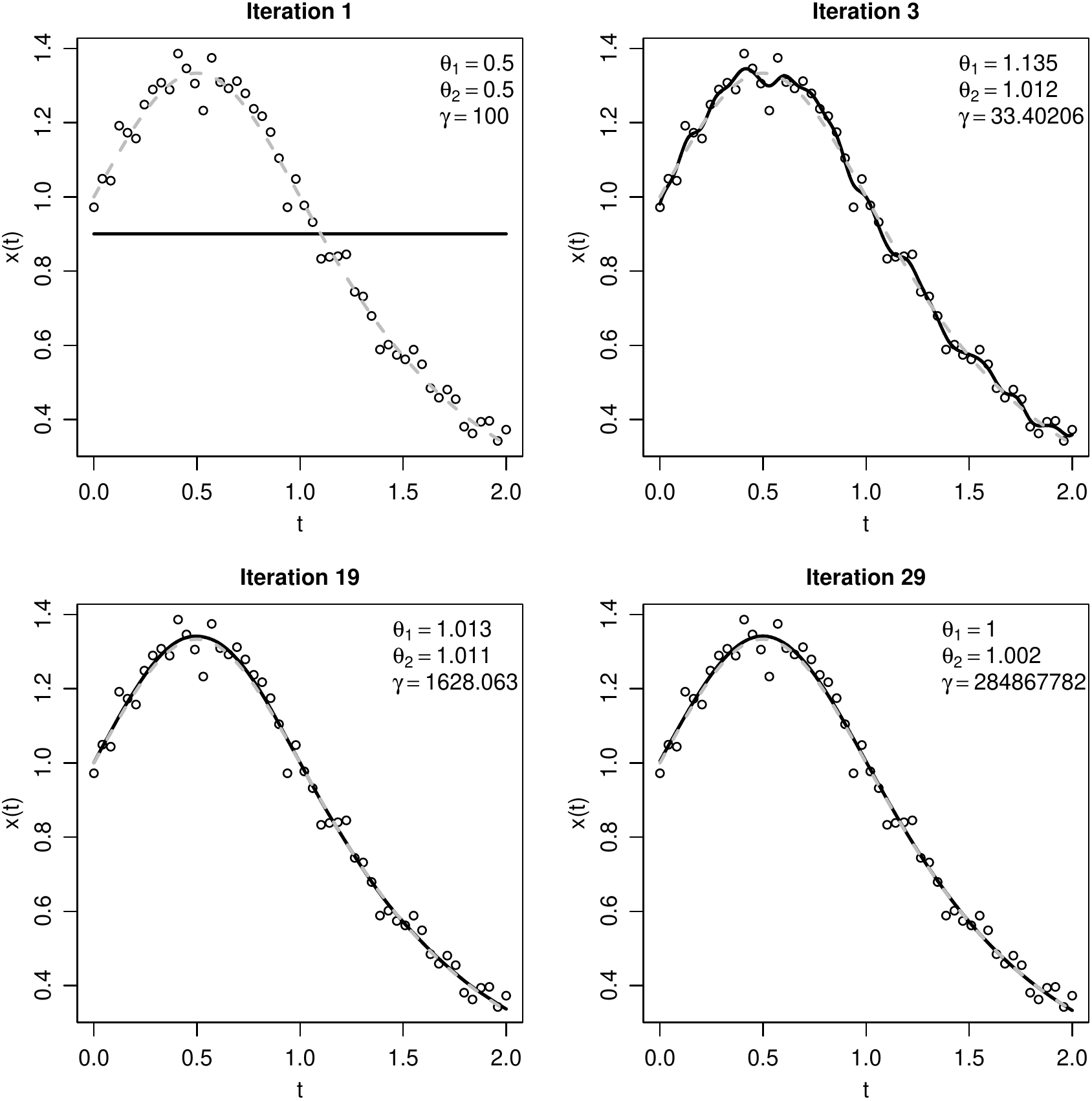}
\caption{Selected sequence of estimation and approximation steps for the first order ODE model in Eq.~(\ref{eq:first_order_example}). Simulated observations are represented by dots while the solid curves correspond to the approximated state functions and the dashed gray ones represent the exact solution to the differential problem. The current estimates of $\boldsymbol{\theta}$ and $\gamma$ are mentioned in the legends.}
\label{fig:robustness_first_order}
\end{figure}

\begin{figure}
\includegraphics[width = 1\textwidth]{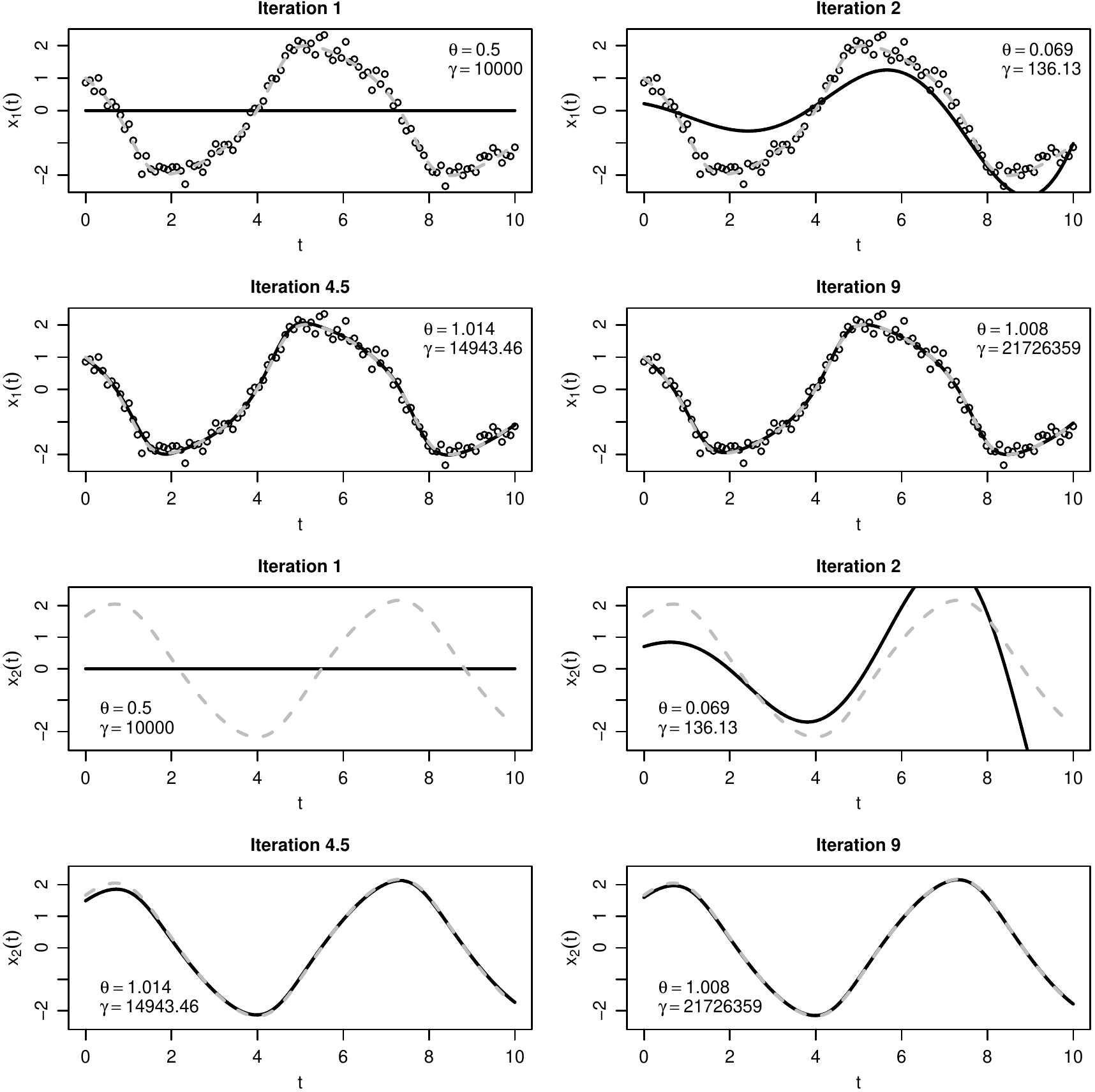}
\caption{Selected sequence of estimation and approximation steps for the Van der Pol system in Eq.~(\ref{eq:first_order_example}). Simulated observations for $x_{1}\left(t\right)$ ($x_{2}\left(t\right)$ is supposed not observed) are displayed by dots while the solid curves correspond to $\widetilde{\boldsymbol{x}}\left(t\right)$, the dashed gray ones represent the numerical solution to the differential problem (obtained using a Runge-Kutta scheme). The current estimates of $\boldsymbol{\theta}$ and $\gamma$ are mentioned in the legends.}
\label{fig:robustness_second_order_x1}
\end{figure}

We propose to linearize the nonlinear part of the ODE model using an iterative approximation procedure. More precisely, given the set of spline coefficients $\boldsymbol{\alpha}^{\left(i\right)}$ at iteration $i$ and the induced estimates of the $j$-th state function $\widetilde{x}_{j}^{\left( i \right)}$, we propose to approximate the $j$-th component of the ODE penalty term at iteration $i + 1$ by:
\begin{equation}
{PEN}_{j}\left(\boldsymbol{\alpha}^{\left(i+1\right)}|\boldsymbol{\alpha}^{\left(i\right)}, \boldsymbol{\theta}\right) \approx \int{\left(\frac{d\widetilde{x}_{j}^{\left(i+1\right)}}{dt}\left(t\right) - f_{j}\left(t, \widetilde{\boldsymbol{x}}^{\left(i\right)}, \boldsymbol{\theta}\right) - \sum_{k = 1}^{d}{\left(\widetilde{x}_{k}^{\left(i+1\right)}\left(t\right) - \widetilde{x}_{k}^{\left(i\right)}\left(t\right)\right) \displaystyle\frac{\partial f_{j}}{\partial x_{k}}\left(t, \widetilde{\boldsymbol{x}}^{\left(i\right)}, \boldsymbol{\theta}\right)}\right)^{2}dt}.
\label{eq:approx_pen_j}
\end{equation}
Then, the overall fidelity measure to the ODE in Eq.~\eqref{eq:global_pen} can be approximated by:
\begin{eqnarray}
{PEN}\left(\boldsymbol{\alpha}^{\left(i+1\right)} | \boldsymbol{\alpha}^{\left(i\right)}, \boldsymbol{\theta}, \boldsymbol{\gamma} \right) & = & \sum_{j = 1}^{d}{\gamma_{j}{PEN}_{j}\left(\boldsymbol{\alpha}^{\left(i+1\right)}|\boldsymbol{\alpha}^{\left(i\right)}, \boldsymbol{\theta}\right)} \nonumber \\
& = & \left(\boldsymbol{\alpha}^{\left(i+1\right)}\right)^{\top}\boldsymbol{R}\left(\boldsymbol{\theta}, \boldsymbol{\gamma},\boldsymbol{\alpha}^{\left(i\right)}\right)\boldsymbol{\alpha}^{\left(i+1\right)} + \left(\boldsymbol{\alpha}^{\left(i+1\right)}\right)^{\top}\boldsymbol{r}\left(\boldsymbol{\theta}, \boldsymbol{\gamma},\boldsymbol{\alpha}^{\left(i\right)}\right) + l\left(\boldsymbol{\theta}, \boldsymbol{\gamma},\boldsymbol{\alpha}^{\left(i\right)}\right),
\label{eq:approx_global_pen}
\end{eqnarray}
where $\boldsymbol{R}\left(\boldsymbol{\theta}, \boldsymbol{\gamma},\boldsymbol{\alpha}^{\left(i\right)}\right)$ plays the role of a penalty matrix, $\boldsymbol{r}\left(\boldsymbol{\theta}, \boldsymbol{\gamma},\boldsymbol{\alpha}^{\left(i\right)}\right)$ is the penalty vector and $l\left(\boldsymbol{\theta}, \boldsymbol{\gamma},\boldsymbol{\alpha}^{\left(i\right)}\right)$ is a constant not depending on the current spline coefficients but only on the previous ones (see Appendix~\ref{section:app_A} for the explicit definition of $\boldsymbol{R}, \boldsymbol{r}$ and $l$). In contrast to the inner optimization proposed in \cite{ramsayetal2007}, the substitution of \eqref{eq:approx_global_pen} in \eqref{eq:goodness_of_fit_profiling} makes the optimization of the criterion $J$ with respect to the spline coefficients a standard least squares problem that does not require any sophisticated mathematical analysis tool. Indeed, Eq.~\eqref{eq:goodness_of_fit_profiling} becomes a quadratic form in $\boldsymbol{\alpha}^{\left(i + 1\right)}$ which optimization does not require the implicit function theorem anymore.  \\
Similarly to \cite{ramsayetal2007} and \cite{frassoetal2013}, the estimation process for the ODE parameters and the spline coefficients implies the iteration of two profiling steps. First, given the last available estimates $\boldsymbol{\alpha}^{\left(i\right)}$ and for given vectors of ODE parameters $\boldsymbol{\theta}^{\left(i\right)}$, precisions of measurements $\boldsymbol{\tau}^{\left(i\right)} = \mbox{vec}\left(\tau^{\left(i\right)}_{j}; j\in\mathcal{J}\right)$ and ODE-compliance parameters $\boldsymbol{\gamma}^{\left(i\right)}$, the spline coefficients $\boldsymbol{\alpha}^{\left(i+1\right)}$ are updated by maximizing the approximated $J$ criterion:
\begin{eqnarray}\label{eq:update_spline_coefficients}
\boldsymbol{\alpha}^{\left(i+1\right)} & = & \mathop{\mbox{argmax}}_{\boldsymbol{\alpha}}J\left(\boldsymbol{\alpha}^{\left(i+1\right)} | \boldsymbol{\theta}^{\left( i \right)}, \boldsymbol{\tau}^{\left( i \right)}, \boldsymbol{\gamma}^{\left( i \right)}, \boldsymbol{\alpha}^{\left(i\right)}, \boldsymbol{y}\right) \nonumber\\
& = & \left(\boldsymbol{G}^{\left( i \right)} + \boldsymbol{R}\left(\boldsymbol{\theta}^{\left( i \right)}, \boldsymbol{\gamma}^{\left( i \right)},\boldsymbol{\alpha}^{\left(i\right)}\right)\right)^{-1}\left(\boldsymbol{g}^{\left( i \right)} - \boldsymbol{r}\left(\boldsymbol{\theta}^{\left( i \right)}, \boldsymbol{\gamma}^{\left( i \right)},\boldsymbol{\alpha}^{\left(i\right)}\right)\right),
\end{eqnarray}
where $\boldsymbol{G}^{\left( i \right)} = \mbox{diag}\left(\boldsymbol{Z}^{\left( i \right)}_{j}; j = 1, \ldots, d\right)$ is a block diagonal matrix with $\boldsymbol{Z}^{\left( i \right)}_{j} = \tau^{\left( i \right)}_{j}\boldsymbol{B}_{j}^{\top}\boldsymbol{B}_{j}$ if the $j$-th state function is observed and the null $K_{j} \times K_{j}$-matrix otherwise and $\boldsymbol{g}^{\left( i \right)} = \mbox{vec}\left(\boldsymbol{z}^{\left( i \right)}_{j}; j = 1, \ldots, d\right)$ with $\boldsymbol{z}^{\left( i \right)}_{j} = \tau^{\left( i \right)}_{j}\boldsymbol{B}_{j}^{\top}\boldsymbol{y}_{j}$ if the $j$-th state function is observed or null ($K_{j}$-vector) otherwise. Here, $\boldsymbol{y}_{j}$ is the $n_{j}$-vector of all observations for the $j$-the state function and $\boldsymbol{B}_{j}$ is a B-spline matrix of dimension $n_{j} \times K_{j}$.
The values of $\boldsymbol{\theta}$ and $\boldsymbol{\tau}$ are updated by optimizing $H$ given $\boldsymbol{\alpha}^{\left( i + 1\right)}$. In the rest of our discussion the iteration indexes for the $\boldsymbol{\theta}$, $\boldsymbol{\gamma}$ and $\boldsymbol{\tau}$ vectors will be omitted. \\
Thanks to the quasilinearization, the selection of the compliance parameters could be performed using standard methods such as AIC, BIC or (generalized) cross-validation. Instead, we advocate to use the EM-type algorithm proposed by \cite{schall1991} which avoids the computation of a selection criterion over a grid of possible values of each $\gamma_{j}$ by estimating them iteratively as the ratio of the estimated error variances and the current variance of the penalty. The standard errors of the ODE parameters can be approximated from the Hessian matrix based on criterion $H$ at convergence. Alternatively, if a quantification including all the sources of uncertainty is desired, we advise to consider the estimation problem from a Bayesian perspective. \\
The following algorithm summarizes the proposed estimation strategy:
\begin{algorithm}
\KwData{Set of (noisy) measurements driven by a dynamic system.}
\KwResult{Estimation of optimal $\boldsymbol{\alpha}$, $\boldsymbol{\theta}$, $\boldsymbol{\gamma}$ and $\boldsymbol{\tau}$ via QL-ODE-P-splines.}
\textbf{Initialization:} Smooth the data using standard P-splines to estimate the initial spline coefficients $\boldsymbol{\alpha}^{(0)}$.
Set the initial $\boldsymbol{\theta}^{(0)}$, $\boldsymbol{\tau}^{(0)}$ and $\boldsymbol{\gamma}^{(0)}$ parameters \;
\While{Convergence is not achieved}{
(\textbf{a}) Given the current $\boldsymbol{\alpha}^{(i)}$ and $\boldsymbol{\gamma}^{(i)}$ linearize the ODE based penalties $PEN_{j}$ (see Web Appendix C) \;

(\textbf{b}) Compute $\boldsymbol{\alpha}^{(i + 1)}$ as Eq.~(\ref{eq:update_spline_coefficients}) using the linearized penalties and compute $\widetilde{x}^{(i+1)}(\boldsymbol{t}) = \boldsymbol{B}\boldsymbol{\alpha}^{(i+1)}$ \;

(\textbf{c}) Update $\boldsymbol{\tau}^{(i + 1)} = \displaystyle \frac{N - \mbox{\textbf{ED}}^{(i+1)}}{\|\boldsymbol{y} - \widetilde{x}^{(i+1)}(\boldsymbol{t}) \|^{2}}$ (the effective dimension is computed as $\mbox{\textbf{ED}}^{(i+1)} = \mbox{tr}\left[\boldsymbol{B} \left(\boldsymbol{G}^{(i)} +  \boldsymbol{R}(\boldsymbol{\theta}^{(i)}, \boldsymbol{\gamma}^{(i)}, \boldsymbol{\alpha}^{(i)}) \right)^{-1} \boldsymbol{\tau}^{(i)}\boldsymbol{B}^{\top}\right] $ and $ \boldsymbol{G}^{(i)} = \boldsymbol{\tau}^{(i)}\boldsymbol{B}^{\top}\boldsymbol{B} $ following \cite{hastieetal1990}) \; 

(\textbf{d}) Compute $\left[\boldsymbol{\sigma}^{2}_{PEN}\right]^{(i+1)} = \displaystyle \frac{PEN\left(\boldsymbol{\alpha}^{(i+1)} | \boldsymbol{\alpha}^{(i)}, \boldsymbol{\theta}^{(i)}, \boldsymbol{\gamma}^{(i)}\right) }{\mbox{\textbf{ED}}^{(i+1)}}$ and ODE compliance parameters $\boldsymbol{\gamma}^{(i+1)} = \displaystyle 1/[\boldsymbol{\sigma}^{2}_{PEN}]^{(i+1)}$\;

(\textbf{e}) Given the current values of the other unknowns, update the ODE parameters by minimizing Eq.~(\ref{eq:ODE_crit})\;

(\textbf{f}) Compute the convergence criterion in Eq.~(\ref{eq:convergence_criterion}).
}
\caption{Quasilinearized ODE-P-spline estimation algorithm.}
\label{algo_ql_ode_p_spline}
\end{algorithm}
The following convergence criterion was used throughout:
\begin{equation}\label{eq:convergence_criterion}
\displaystyle \left[\max \left( \left| \frac{\boldsymbol{\alpha}^{(i + 1)} - \boldsymbol{\alpha}^{(i)}}{\boldsymbol{\alpha}^{(i)}} \right| \right), \max\left( \left| \frac{\boldsymbol{\gamma}^{(i + 1)} - \boldsymbol{\gamma}^{(i)}}{\boldsymbol{\gamma}^{(i)}} \right| \right), \max\left( \left| \frac{\boldsymbol{\theta}^{(i + 1)} - \boldsymbol{\theta}^{(i)}}{\boldsymbol{\theta}^{(i)}} \right| \right) , \max\left( \left| \frac{\boldsymbol{\tau}^{(i + 1)} - \boldsymbol{\tau}^{(i)}}{\boldsymbol{\tau}^{(i)}} \right| \right)\right] < 10^{-4}.
\end{equation}
The initial values for the ODE parameters and the precision of measurements can be either chosen by the analyst on the basis of prior knowledge or estimated from the raw measurements. In the latter case different alternatives are possible. One possibility is to invert the ODE system using the spline coefficients estimated by P-splines. Alternatively, the estimates of a NLS analysis (based on the numerical solution of the ODE) can be used. The ODE compliance parameters can be set at large values (as in \cite{ramsayetal2007}) if the interest is exclusively focused on ODE parameter estimation. Then, step (d) in Algorithm~\ref{algo_ql_ode_p_spline} can be skipped. \\
The proposed procedure requires an initial vector of spline coefficients: we recommend to use a standard P-spline smoother \citep{eilersandmarx1996} to start with reasonable initial coefficients $\boldsymbol{\alpha}^{(0)}$. The ODE model combined with the initial ODE parameters ($\boldsymbol{\theta}^{(0)}$) and the initial spline coefficients associated to the observed state functions can be used to deduce initial spline coefficients for the unobserved state functions. On the other hand, according to our experience, the quality of the final estimates is not dramatically influenced by the choice of $\boldsymbol{\alpha}^{(0)}$. \\
This issue is illustrated in the analysis of the results (Table~\ref{tab:robustness_ODEs}) of a simulation study. One hundred data sets have been generated by adding a zero mean Gaussian noise to the solutions of the ODE problems presented in Section~\ref{section:introduction}. These data have been analyzed using QL-ODE-P-splines with initial ODE parameters set either to their true values $\boldsymbol{\theta}^{*}$, to $0.5\boldsymbol{\theta}^{*}$ or to $2 \boldsymbol{\theta}^{*}$. The initial spline coefficients have been obtained: by smoothing of the raw measurements with standard P-splines (Strategy 1), or by assuming constant state functions (Strategy 2). The smoothers have been built on cubic B-splines defined over 20 equidistant internal knots to approximate the state function in the first order ODE model (Example 1) and using fourth order spline functions computed on 150 equidistant internal knots for the simulation of the Van der Pol system (Example 2). The respective sample sizes are $N = 50$ and $100$.\\
Figure~\ref{fig:robustness_first_order} shows possible estimation steps for Example 1 starting with a constant state function and wrong initial ODE parameters. These plots clearly show the impact of the bad initial guesses (upper left panel). At the second estimation step, the ODE-compliance parameter $\gamma$ becomes small giving more weight to the goodness-of-fit term in Eq.~\eqref{eq:goodness_of_fit_profiling}. Then, the (fitted) red curve tends to overfit the data with estimated ODE parameters different from the true ones ($\boldsymbol{\theta}^{*}_{1} = \boldsymbol{\theta}^{*}_{2} = 1$). As the number of iterations increases, the ODE compliance parameter also increases, yielding more weight to the ODE-based penalty in Eq.~\eqref{eq:goodness_of_fit_profiling} and ensuring estimated state functions and ODE parameters closer and closer to their true values.
Figures~\ref{fig:robustness_second_order_x1} shows similar results for Example 2 (based on the Van der Pol system). It is remarkable that, even with initial spline coefficients chosen without taking into account the ODE model and/or the observations, we properly approximate both the observed ($x_{1}(t)$) and unobserved ($x_{2}(t)$) state functions. It suggests that results provided by Algorithm~\ref{algo_ql_ode_p_spline} are robust to the choice of initial conditions.\\
Table~\ref{tab:robustness_ODEs} shows the (average) root mean squared error w.r.t. the data ($\overline{RMSE(\boldsymbol{y})} = 0.01 \sum_{k=1}^{100} \sqrt{N^{-1} \|\boldsymbol{y}_{k} - \boldsymbol{\widetilde{x}}_{k} \|^2 }$) and the (average) RMSE w.r.t. the state function ($\overline{RMSE(\boldsymbol{x})} = 0.01 \sum_{k=1}^{100} \sqrt{N^{-1} \|\boldsymbol{x}_{k} - \boldsymbol{\widetilde{x}}_{k}\|^2} $): the influence of the initial guesses on the goodness of the final fit appears negligible. Analogously, the compliance of the approximated state functions to the simulated ones does not seem affected by the specification of the initial settings. On average, the choice of the initial values for the spline coefficients seems to have a limited impact on the number of iterations required to achieve converge ($\overline{\mbox{\# Iter}}$). 

\begin{table}[htbp]
  \centering
\caption{Robustness of the QL-ODE-P-spline estimation approach to the choice of initial conditions.
The first two rows of the table show the (average) RMSEs between the approximated state function and the simulated measurements taking either identical initial values for the spline coefficients (Strategy 1) or a standard P-spline smoother (Strategy 2). Rows 3 and 4 report the (average) RMSEs computed with respect to the analytic/numerical solution of the ODE model. The remaining part of the table reports the average number of iterations ($\overline{\mbox{\# Iter}}$) required to achieve convergence.
}
\begin{tabular}{cc ccc c ccc}
      & \multicolumn{1}{r}{} & \multicolumn{3}{c}{\textbf{First order ODE}} & & \multicolumn{3}{c}{\textbf{Van der Pol system}} \\
\cline{2-9}      & Initial $\boldsymbol{\alpha}$ & $\boldsymbol{\theta}^{*}$ & $0.5 \boldsymbol{\theta}^{*}$ & $2\boldsymbol{\theta}^{*}$ & & $\boldsymbol{\theta}^{*}$ & $0.5 \boldsymbol{\theta}^{*}$ & $2\boldsymbol{\theta}^{*}$ \\
\cline{1-9}
\multirow{2}[1]{*}{$\overline{RMSE(\boldsymbol{y})}$} & Strategy 1 & 0.041 & 0.040 & 0.041&  & 0.042 & 0.040 & 0.041 \\																							
\multicolumn{1}{c}{} & Strategy 2 & 0.042 & 0.040 & 0.041 & & 0.042 & 0.040 & 0.041 \\[0.2cm]
\multirow{2}[1]{*}{$\overline{RMSE(\boldsymbol{x})}$} & Strategy 1 & 0.012 & 0.013 & 0.013 & & 0.011 & 0.013 & 0.013 \\
      & Strategy 2 & 0.011 & 0.013 & 0.013 & & 0.011 & 0.013 & 0.013\\[0.2cm]
\multirow{2}[1]{*}{$\overline{\mbox{\# Iter}}$}		&  Strategy 1 	 & 16 & 22 & 21  & & 16 & 14 &  14		\\			
		&  Strategy 2	 & 18 & 23 & 21 &	& 16 & 14 & 14   			 \\
\cline{1-9}
\end{tabular}%
  \label{tab:robustness_ODEs}%
\end{table}%
\subsection{Including state conditions in the estimation process}
\label{subsection:differential_conditions}
A (system of) differential equation(s) can have a unique solution if condition(s) are imposed on the value(s) of the state function(s) and/or of one of its derivatives at a given time point(s). Initial and boundary value conditions are usually distinguished. In what follows, for the sake of brevity, we generically refer to them as ``state conditions''.\\
Often state conditions arise as characteristic of the observed dynamics. Such information can easily be included in the estimation process with a linear differential model \citep{frassoetal2013}. The same is true after quasilinearization of a nonlinear system. This is an extra advantage over exiting approaches in the literature \citep{ramsayetal2007, caoandzhao2008, poytonetal2006}.\\
Consider the general differential problem given by:
\begin{equation*}
\left\{\begin{array}{rcl}
       \displaystyle\frac{d\boldsymbol{x}}{dt}\left(t\right) & = &\boldsymbol{f}\left(t, \boldsymbol{x}, \boldsymbol{\theta}\right), \\
       \displaystyle \boldsymbol{x}\left(t_{0}\right) & = & \boldsymbol{s}_{t_{0}}\ \mbox{for}\ t_{0} \in \mathcal{T}_{0},
       \end{array}
\right.
\label{eq:general_def_ode_boundaries}
\end{equation*}
where $\mathcal{T}_{0}$ denotes the subset of observed times where $ \boldsymbol{x}$ is forced to be equal to $\boldsymbol{s}_{t_{0}}$. Using the B-spline approximation, these conditions become:
\begin{equation}
\boldsymbol{S}\boldsymbol{\alpha}^{\left(i+1 \right)} = \boldsymbol{s}_{t_{0}},
\label{eq:b_spline_representation_boundaries}
\end{equation}
where $\boldsymbol{S}$ is a matrix where each of the rows contains the B-splines in the basis function evaluated at a given $t_{0} \in \mathcal{T}_{0}$. These state conditions can be included in the estimation process using two strategies.\\
The first one consists in treating them as an extra least squares penalty. Then, the optimal spline coefficient can be obtained by maximizing the following modified fitting criterion:
\begin{equation}\label{eq:goodness_of_fit_ls_extra_pen}
\begin{array}{r c l}
J\left(\boldsymbol{\alpha}^{\left(i+1\right)}|\boldsymbol{\theta}, \boldsymbol{\gamma}, \boldsymbol{\tau}, \boldsymbol{\alpha}^{\left(i\right)}, \boldsymbol{y}\right) & = & \displaystyle\sum_{j\in \mathcal{J}}{\left\{\displaystyle\frac{n_{j}}{2}\log\left(\tau_{j}\right) - \displaystyle\frac{\tau_{j}}{2}\left\|\boldsymbol{y}_{j} - \widetilde{x}_{j}^{\left(i+1\right)}\left(\boldsymbol{t}\right)\right\|^{2}\right\}} - \displaystyle\frac{1}{2}{PEN}\left(\boldsymbol{\alpha}^{\left(i+1\right)}|\boldsymbol{\alpha}^{\left(i\right)}, \boldsymbol{\gamma}, \boldsymbol{\theta} \right)\\
&& - \displaystyle\frac{\kappa}{2}\left(\boldsymbol{S}\boldsymbol{\alpha}^{\left(i+1\right)} - \boldsymbol{s}_{t_{0}}\right)^{\top}\left(\boldsymbol{S}\boldsymbol{\alpha}^{\left(i+1\right)} - \boldsymbol{s}_{t_{0}}\right),
\end{array}
\end{equation}
where $\kappa$ is set at an arbitrary large value. Note that with such an extra penalty, the state condition is imposed only in a least square sense. With the fitting criterion in Eq.~\eqref{eq:goodness_of_fit_ls_extra_pen}, the optimal spline coefficients at iteration $(i + 1)$ become:
\begin{eqnarray*}
\boldsymbol{\alpha}^{\left(i+1\right)} & = & \mathop{\mbox{argmax}}_{\boldsymbol{\alpha}}J\left(\boldsymbol{\alpha}^{\left(i+1\right)} | \boldsymbol{\theta}, \boldsymbol{\tau}, \boldsymbol{\gamma}, \boldsymbol{\alpha}^{\left(i\right)}\boldsymbol{y}\right) \nonumber\\
& = & \left(\boldsymbol{G} + \boldsymbol{R}\left(\boldsymbol{\theta}, \boldsymbol{\gamma},\boldsymbol{\alpha}^{\left(i\right)}\right) + \kappa \boldsymbol{S}^{\top}\boldsymbol{S}\right)^{-1}\left(\boldsymbol{g} - \boldsymbol{r}\left(\boldsymbol{\theta}, \boldsymbol{\gamma},\boldsymbol{\alpha}^{\left(i\right)}\right) + \kappa \boldsymbol{S}^{\top}\boldsymbol{s}_{t_{0}}\right).
\end{eqnarray*}
If $\kappa$ is equal to zero, then one gets back the estimation procedure discussed before, while for $\kappa$ tending to infinity, the state functions are forced to meet the imposed conditions. This parameter can be fixed to a large value ($\kappa = 10^{6}$, say) if the user is confident in the validity of the conditions or optimized using standard criteria (such AIC, BIC, cross-validation). \\
The second approach consists in forcing the smooth estimates of the state function to check the prescribed conditions by the introduction of Lagrange multipliers. Starting from Eq.~\eqref{eq:goodness_of_fit_profiling} the Lagrange function for the constrained optimization problem is:
\begin{equation*}
\begin{array}{r c l}
\mathcal{L}\left(\boldsymbol{\alpha}^{\left(i+1\right)}, \boldsymbol{\omega}|\boldsymbol{\theta}, \boldsymbol{\gamma}, \boldsymbol{\tau}, \boldsymbol{\alpha}^{\left(i\right)}, \boldsymbol{y}\right) & = & \displaystyle\sum_{j\in \mathcal{J}}{\left\{\displaystyle\frac{n_{j}}{2}\log\left(\tau_{j}\right) - \displaystyle\frac{\tau_{j}}{2}\left\|\boldsymbol{y}_{j} - \widetilde{x}_{j}^{\left(i+1\right)}\left(\boldsymbol{t}\right)\right\|^{2}\right\}} - \displaystyle\frac{1}{2}{PEN}\left(\boldsymbol{\alpha}^{\left(i+1\right)}|\boldsymbol{\alpha}^{\left(i\right)}, \boldsymbol{\gamma}, \boldsymbol{\theta}\right)\\
&& - \boldsymbol{\omega}^{\top}\left(\boldsymbol{S}\boldsymbol{\alpha}^{\left(i+1\right)} - \boldsymbol{s}_{t_{0}}\right),
\end{array}
\end{equation*}
where $\boldsymbol{\omega}$ is the vector of Lagrange multipliers. Following \cite{currie2013}, the maximization of $\mathcal{L}$ can be obtained by solving:
\begin{equation*}
\left(\begin{array}{c c} \boldsymbol{G} + \boldsymbol{R}\left(\boldsymbol{\theta}, \boldsymbol{\gamma},\boldsymbol{\alpha}^{\left(i\right)}\right) & \boldsymbol{S}^{\top} \\
                         \boldsymbol{S} & \boldsymbol{0}
      \end{array}
\right)
\left(\begin{array}{c} \boldsymbol{\alpha}^{\left(i+1\right)} \\
                       \boldsymbol{\omega}
      \end{array}
\right) =
\left(\begin{array}{c} \boldsymbol{g} - \boldsymbol{r}\left(\boldsymbol{\theta}, \boldsymbol{\gamma},\boldsymbol{\alpha}^{\left(i\right)}\right) \\
                       \boldsymbol{s}_{t_{0}}
      \end{array}
\right).
\end{equation*}
Whatever the selected approach to force the state conditions, the estimates of the ODE parameters $\boldsymbol{\theta}$ are updated using Eq.~\eqref{eq:ODE_crit}.
\section{Practical implementation and simulation}
\label{section:simulation}
In this section the results of simulation studies based on the two examples presented in Section~\ref{section:introduction} are reported to illustrate the properties of the proposed approach. It is first shown how these ODEs can be linearized to define the steps in the QL-ODE-P-spline estimation algorithm. \\
The ODE given in Eq.~\eqref{eq:first_order_example} can be linearized at iteration $i + 1$ as follows:
\begin{equation*}
\begin{array}{r c l}
\displaystyle \frac{d \widetilde{x}^{\left( i + 1 \right)}}{d t}(t) &= &(\theta_{1} - 2 \theta_{2} t) \left[\widetilde{x}^{\left( i + 1 \right)} \left(t\right) \right]^{2}\\                                               
                                                  &\approx & 2 \left(\theta_{1} - 2 \theta_{2} t\right)\widetilde{x}^{\left(i\right)}\left(t\right)\widetilde{x}^{\left(i+1\right)}\left(t\right) - (\theta_{1} - 2 \theta_{2} t) \left[\widetilde{x}^{\left( i \right)}(t)\right]^{2}.
\end{array}
\end{equation*}
In the Van der Pol system, only the first differential equation has a nonlinear term. It can be linearized using:
\begin{equation*}
\begin{array}{r c l}
\displaystyle{\frac{d \widetilde{x}_{1}^{\left(i + 1 \right)} } {d t}} \left(t\right) & = & \theta \left(\widetilde{x}_{1}^{\left(i + 1\right)}\left(t\right) - \displaystyle{ \frac{1}{3}} \left[\widetilde{x}_{1}^{\left(i + 1\right)} \left(t \right)\right]^{3} - \widetilde{x}_{2}^{\left(i + 1 \right)}\left(t\right)\right)\\
&\approx& \theta \left(1  - \left[\widetilde{x}_{1}^{\left(i\right)} \left(t \right) \right]^{2} \right) \widetilde{x}_{1}^{\left(i + 1 \right)}\left(t \right) - \theta \widetilde{x}_{2}^{\left(i + 1\right)}\left(t \right) + \displaystyle{\frac{2\theta}{3}} \left[\widetilde{x}_{1}^{\left(i\right)} \left(t \right) \right]^{3}.
 \end{array}
\end{equation*}
These approximations are then used to build the penalty terms in Eqs.~\eqref{eq:pen_j} and~\eqref{eq:global_pen}. The spline coefficients and the other parameters are jointly updated with the ODE-compliance parameters (using the EM-type approach by \cite{schall1991}). The QL-ODE-P-spline estimation steps are summarized in Algorithm~\ref{algo_ql_ode_p_spline}.
\subsection{Evaluation of estimation accuracy}
\label{subsection:estimation_accuracy}
An intensive simulation study is presented here to study the properties of the proposed methods. We have generated a large number of datasets using the ODE models in Eqs.~\eqref{eq:first_order_example} and~\eqref{eq:second_order_example} by perturbing the analytic or numerical solutions of the differential models by a zero mean Gaussian noise. \\
We aim to assess the quality of the estimates provided by the QL-ODE-P-spline approach for known or unknown initial values for the state functions. These constraints were imposed during QL-ODE-P-spline estimation using Lagrange multipliers (see Section~\ref{subsection:differential_conditions}).
\subsubsection{First order ODE simulation results}
\label{subsubsection:first_order_estimation_accuracy}
In this section, the results of simulations based on Eq.~\eqref{eq:first_order_example} are discussed. The (analytic) state function has been computed with $\theta_{1} = \theta_{2} = 1$ and $x(0) = 1$. Gaussian noise with mean zero and variance $\sigma^{2} = 0.045^{2}$ has been added to $N$ values of the state function evaluated at time points uniformly sampled in $[0, 2]$. The data generation process has been repeated 500 times under four sample sizes $N = \left(20, 50, 100, 500 \right)$.\\
Table~\ref{tab:simulation_first_order_no_conditions} summarizes the estimation performances when $x(0)$ is unknown and provides the bias (in percentage), the averaged standard errors, the relative mean squared error and standard deviation of $\widehat{\theta}_{1}, \widehat{\theta}_{2}$ and $\widehat{\tau}$. In all the simulation settings we used cubic B-splines defined on 20 equidistant internal knots. The quality of the QL-ODE-P-spline estimates are compared with those obtained by estimating $\boldsymbol{\theta}$ from Eq.~\eqref{eq:solution_first_order} using nonlinear least squares.\\
The table clearly shows that the NLS and the quasilinearized ODE-P-spline approaches provide good estimates of the unknown ODE parameters. The biases for all the parameter estimates are small and similar for the two procedures whatever the sample size. Similar results are obtained with the two methods for the estimation of the standard deviations and the root mean squared errors of the parameter estimators.\\
\begin{table}[htbp]
  \centering
  \caption{Simulation results for the first order ODE example without initial conditions. For the two approaches (quasilinearized ODE-P-spline and NLS), the bias, the root mean squared errors and the standard deviations of the estimated parameters (ODE and initial values) are provided. In addition for each ODE parameter the averaged standard errors are reported. The vector of (averaged) compliance parameters optimized for the spline based approach is indicated in the last row of the table. }
\begin{tabular}{rrrrrrrrrr}
      & \multicolumn{1}{r}{} & \multicolumn{4}{c}{\textbf{NLS estimates}} & \multicolumn{4}{c}{\textbf{QL-ODE-P-spline estimates}}\\
\hline
      & \multicolumn{1}{r}{} & $N = 20$ & $N = 50$ & $N = 100$ & $N = 500$ & $N = 20$ & $N = 50$ & $N = 100$ & $N = 500$ \\
\hline
      & \multicolumn{1}{r}{} &       &       &       &       &       &       &       & \\
\multicolumn{1}{r}{\multirow{4}[0]{*}{\textbf{Bias}}} & $\theta_{1}$ & -0.242\% & 0.038\% & 0.004\% & -0.057\% & -0.242\% & 0.038\% & 0.004\% & -0.058\% \\
\multicolumn{1}{c}{} & $\theta_{2}$ & -0.176\% & 0.022\% & 0.018\% & -0.052\% & -0.182\% & 0.024\% & -0.178\% & -0.047\% \\
\multicolumn{1}{c}{} & $x(0)$ & 0.112\% & 0.034\% & 0.025\% & 0.018\% & 0.082\% & 0.024\% & 0.019\% & 0.017\% \\
\multicolumn{1}{c}{} & $\tau$ & 18.29\% & 4.83\% & 2.61\% & 0.47\% & 32.20\% & 9.29\% & 4.72\% & 0.89\% \\
      & \multicolumn{1}{r}{} &       &       &       &       &       &       &       & \\
\multicolumn{1}{r}{\multirow{3}[0]{*}{\textbf{Mean s.e.}}} & $\theta_{1}$ & 7.9E-02 & 5.3E-02 & 3.8E-02 & 1.7E-02 & 7.5E-02 & 5.2E-02 & 3.8E-02 & 1.7E-02\\ 
\multicolumn{1}{c}{} & $\theta_{2}$ & 5.7E-02 & 3.8E-02 & 2.7E-02 & 1.2E-02 & 5.4E-02 & 3.7E-02 & 2.7E-02 & 1.2E-02\\ 
\multicolumn{1}{c}{} & $x(0)$ & 2.5E-02 & 1.7E-02 & 1.2E-02 & 5.5E-03 &       &       &       &  \\
      & \multicolumn{1}{r}{} &       &       &       &       &       &       &       &  \\
\multicolumn{1}{r}{\multirow{4}[0]{*}{\textbf{RMSE}}} & $\theta_{1}$ & \textless 1.0E-03 & \textless 1.0E-03 & \textless 1.0E-03 & \textless 1.0E-03 & 3.1E-03 & 4.8E-03 & \textless1.0E-03 & \textless1.0E-03 \\
\multicolumn{1}{c}{} & $\theta_{2}$ & \textless 1.0E-03  & \textless 1.0E-03  & \textless 1.0E-03  & \textless 1.0E-03  & 9.3E-03 & 2.6E-03 & \textless 1.0E-03  & \textless 1.0E-03  \\
\multicolumn{1}{c}{} & $x(0)$ & \textless 1.0E-03 & \textless 1.0E-03 & \textless 1.0E-03 & \textless 1.0E-03 & \textless 1.0E-03 & \textless 1.0E-03 & \textless 1.0E-03 & \textless 1.0E-03 \\
\multicolumn{1}{c}{} & $\tau$ & 2.5E-01 & 5.9E-02 & 6.7E-02 & 4.5E-03 & 1.8E-02 & 8.1E-03 & 1.4E-02 & \textless 1.0E-03 \\
      & \multicolumn{1}{r}{} &       &       &       &       &       &       &       &  \\
\multicolumn{1}{r}{\multirow{4}[1]{*}{\textbf{Std dev}}} & $\theta_{1}$ & 7.9E-02 & 5.5E-02 & 3.6E-02 & 1.7E-02 & 7.9E-02 & 5.5E-02 & 3.6E-02 & 1.7E-02 \\
\multicolumn{1}{c}{} & $\theta_{2}$ & 5.6E-02 & 3.9E-02 & 2.5E-02 & 1.2E-02 & 5.7E-02 & 3.9E-02 & 2.5E-02 & 1.2E-02 \\
\multicolumn{1}{c}{} & $x(0)$ & 2.5E-02 & 1.8E-02 & 1.2E-02 & 5.5E-03 & 2.5E-02 & 1.8E-02 & 1.2E-02 & 5.5E-03 \\
\multicolumn{1}{c}{} & $\tau$ & 4.9E-01 & 2.2E-01 & 1.5E-01 & 6.6E-02 & 2.5E-02 & 2.3E-01 & 1.5E-01 & 6.7E-02\\
      & \multicolumn{1}{r}{} &       &       &       &       &       &       &       &  \\
\multicolumn{1}{c}{} & $\bar{\gamma}$ &  &  &  &  & 2.5E+07	& 2.4E+07	& 2.4E+07	& 2.4E+07\\
      & \multicolumn{1}{r}{} &       &       &       &       &       &       &       & \\
\hline
\end{tabular}%
  \label{tab:simulation_first_order_no_conditions}%
\end{table}%
From Table~\ref{tab:simulation_first_order_conditions} one can evaluate the impact of the initial value conditions on the estimation performances. These constraints reduce the estimation biases of the two procedures whatever the sample size. Also in this case the standard deviations and the root mean squared errors of the parameter estimators appear really similar for the two alternative procedures. 
\begin{table}[htbp]
  \centering
  \caption{Simulation results for the first order ODE example with known initial conditions. For the quasilinearized ODE-P-spline approach the initial conditions have been introduced using Lagrange multipliers (Section~\ref{subsection:differential_conditions}). For the two approaches, the bias, the root mean squared errors and the standard deviations computed for the estimates of the parameters are shown. In addition for the ODE parameters the average standard errors estimated over the simulation runs are reported. For the QL-ODE-P-spline the vector of (averaged) optimal compliance parameters is indicated in the last row of the table. }
	\begin{tabular}{rrrrrrrrrr}
      & \multicolumn{1}{c}{} & \multicolumn{4}{c}{\textbf{NLS estimates}} & \multicolumn{4}{c}{\textbf{QL-ODE-P-spline estimates}} \\
\hline
      & \multicolumn{1}{r}{} & $N = 20$ & $N = 50$ & $N = 100$ & $N = 500$ & $N = 20$ & $N = 50$ & $N = 100$ & $N = 500$ \\
\hline
      & \multicolumn{1}{r}{} &       &       &       &       &       &       &       &  \\
\multicolumn{1}{r}{\multirow{3}[0]{*}{\textbf{Bias}}} & $\theta_{1}$ &-0.098\%	&0.033\%	&0.048\%	&-0.009\%	&-0.085\%	&0.036\%&	0.159\% &	0.036\% \\
\multicolumn{1}{c}{} & $\theta_{2}$ &-0.095\%	&0.017\%&	0.024\%	&-0.020\%	&-0.085\%	&0.036\% &	0.159\% &	0.036\% \\
\multicolumn{1}{c}{} & $\tau$ & 18.05\% & 4.61\% & 2.63\% & 0.48\% & 11.11\% & 2.54\% & 1.69\% & 0.32\% \\
      & \multicolumn{1}{c}{} &       &       &       &       &       &       &       &  \\
\multicolumn{1}{r}{\multirow{2}[0]{*}{\textbf{Mean s.e.}}} & $\theta_{1}$ & 3.8E-02 & 2.4E-02 & 1.7E-02 & 7.7E-03 & 3.9E-02 & 2.5E-02 & 2.1E-02 & 1.4E-02 \\
\multicolumn{1}{c}{} & $\theta_{2}$ & 4.0E-02 & 2.5E-02 & 1.8E-02 & 8.1E-03 & 4.1E-02 & 2.6E-02 & 2.1E-02 & 1.2E-02 \\
      & \multicolumn{1}{r}{} &       &       &       & \multicolumn{1}{r}{} &       &       &       &  \\
\multicolumn{1}{r}{\multirow{3}[0]{*}{\textbf{RMSE}}} & $\theta_{1}$ & 3.7E-03 & 3.5E-03 & 5.5E-04 & \textless 1.0E-03 & 3.5E-03 & 3.5E-03 & \textless 1.0E-03 & \textless 1.0E-03  \\
\multicolumn{1}{c}{} & $\theta_{2}$ & 9.8E-03 & 2.1E-03 & \textless 1.0E-03  & \textless 1.0E-03  & 3.5E-03 & 3.5E-03 & \textless 1.0E-03  & \textless 1.0E-03  \\
\multicolumn{1}{c}{} & $\tau$ & 3.3E-02 & 4.4E-03 & 6.7E-03 & \textless 1.0E-03 & 1.5E-01 & \textless 1.0E-03  & 5.6E-03 & \textless 1.0E-03  \\
      & \multicolumn{1}{r}{} &       &       &       & \multicolumn{1}{r}{} &       &       &       &  \\
\multicolumn{1}{r}{\multirow{3}[0]{*}{\textbf{Std dev}}} & $\theta_{1}$ & 4.0E-02 & 2.5E-02 & 1.7E-02 & 7.2E-03 & 4.0E-02 & 2.5E-02 & 3.0E-02 & 2.2E-02 \\
\multicolumn{1}{c}{} & $\theta_{2}$ & 4.2E-02 & 2.6E-02 & 1.8E-02 & 7.7E-03 & 4.0E-02 & 2.5E-02 & 3.0E-02 & 2.2E-02 \\
\multicolumn{1}{c}{} & $\tau$ & 4.8E-01 & 2.2E-01 & 1.5E-01 & 6.6E-02 & 4.4E-01 & 2.1E-01 & 1.5E-01 & 6.6E-02 \\
      & \multicolumn{1}{r}{} &       &       &       & \multicolumn{1}{r}{} &       &       &       &  \\
      & $\bar{\gamma}$ &       &       &       &       & 5.0E+07 & 4.8E+07 & 4.8E+07 & 4.7E+07 \\
      & \multicolumn{1}{c}{} &       &       &       & \multicolumn{1}{c}{} &       &       &       &  \\
\hline
\end{tabular}%
 \label{tab:simulation_first_order_conditions}%
\end{table}%
\subsubsection{Van der Pol system simulation results}
\label{subsubsection:second_order_estimation_accuracy}
A simulation study was also performed with the Van der Pol system in Eq.~\eqref{eq:second_order_example}. Five hundreds datasets of size $N = (50, 100, 500, 1000)$ have been obtained adding a Gaussian noise component with null mean and variance $\tau^{-1} = \sigma^2 = 0.07^{2}$ to uniformly sampled values of the numerical solution to Eq.~\eqref{eq:second_order_example}. The solution has been approximated using a Runge-Kutta scheme for values of $t$ in $[0, 10]$ and considering initial value conditions $x_{1}(0) = x_{2}(0) = 1$ for $\theta = 1$. Table~\ref{tab:simulation_second_order} compares constrained and unconstrained QL-ODE-P-spline methods in terms of estimation performances. These results have been obtained using fourth order B-splines defined on 150 equidistant internal knots.\\
\begin{table}[htbp]
  \centering
  \caption{Simulation results with a Van der Pol system: the biases, root mean squared errors and standard deviations for parameter estimates using the QL-ODE-P-spline approach (with and without state conditions). In the last two rows of the upper and lower parts of the table, the (average) optimal ODE-compliance parameters are shown together with the (average) ODE parameter standard errors.}
\begin{tabular}{rrrrrrr}
      &       &       & $N = 50$ & $N = 100$ & $N = 500$ & $N = 1000$ \\
\hline
      &       &       &       &       &       & \\
\multicolumn{1}{c}{\multirow{16}[0]{*}{\begin{sideways}\textbf{Unknown differential conditions}\end{sideways}}} & \multicolumn{1}{c}{\multirow{4}[0]{*}{\textbf{Bias}}} & $\theta = 1$ & 0.165\% & -0.015\% & -0.002\% & -0.009\% \\
\multicolumn{1}{c}{} & \multicolumn{1}{c}{} & $x_{1}(0) = 1$ & -0.178\% & -0.045\% & 0.015\% & -0.021\% \\
\multicolumn{1}{c}{} & \multicolumn{1}{c}{} & $x_{2}(0) = 1$ & -0.166\% & 0.059\% & 0.034\% & -0.002\% \\
\multicolumn{1}{c}{} & \multicolumn{1}{c}{} & $\tau = 200$ & 7.495\% & 3.936\% & 0.699\% & 0.292\% \\
\multicolumn{1}{c}{} &       &       &       &       &       &  \\
\multicolumn{1}{c}{} & \multicolumn{1}{c}{\multirow{4}[0]{*}{\textbf{RMSE}}} & $\theta $ &  \textless 1.0E-03 &  \textless 1.0E-03 &  \textless 1.0E-03 &  \textless 1.0E-03\\
\multicolumn{1}{c}{} & \multicolumn{1}{c}{} & $x_{1}(0) $ &  \textless 1.0E-03 &  \textless 1.0E-03 &  \textless 1.0E-03 &  \textless 1.0E-03 \\
\multicolumn{1}{c}{} & \multicolumn{1}{c}{} & $x_{2}(0)$ &  \textless 1.0E-03 &  \textless 1.0E-03 &  \textless 1.0E-03 &  \textless 1.0E-03\\
\multicolumn{1}{c}{} & \multicolumn{1}{c}{} & $\tau $ & 3.8E+00 & 3.7E+00 & 3.7E-02 & 1.2E-01 \\
\multicolumn{1}{c}{} &       &       &       &       &       &  \\
\multicolumn{1}{c}{} & \multicolumn{1}{c}{\multirow{4}[0]{*}{\textbf{Std dev}}} & $\theta $ & 3.3E-02 & 2.4E-02 & 1.0E-02 & 8.3E-03 \\
\multicolumn{1}{c}{} & \multicolumn{1}{c}{} & $x_{1}(0) $ & 3.1E-02 & 2.2E-02 & 9.2E-03 & 7.0E-03 \\
\multicolumn{1}{c}{} & \multicolumn{1}{c}{} & $x_{2}(0)$ & 5.8E-02 & 4.1E-02 & 1.8E-02 & 1.4E-02 \\
\multicolumn{1}{c}{} & \multicolumn{1}{c}{} & $\tau $ & 2.3E-01 & 1.E-01 & 6.6E-02 & 4.6E-02 \\
\multicolumn{1}{c}{} & \multicolumn{1}{c}{} &       &       &       &       &  \\
\multicolumn{1}{c}{} &       & $\overline{se(\theta)}$  & 2.78E-02 & 2.05E-02 & 9.48E-03 & 6.74E-03 \\
\multicolumn{1}{c}{\textbf{}} &       & $\bar{\gamma}$  & 8.7E+07 & 8.7E+07 & 8.7E+07 & 8.6E+07 \\
\multicolumn{1}{c}{\textbf{}} &       &       &       &       &       &  \\
\hline
      &       &       &       &       &       &  \\
\multicolumn{1}{c}{\multirow{10}[0]{*}{\begin{sideways}\textbf{Known differential conditions}\end{sideways}}} & \multicolumn{1}{c}{\multirow{2}[0]{*}{\textbf{Bias}}} & $\theta = 1$ & 0.037\% & 0.02\% & 0.10\% & -0.02\% \\
\multicolumn{1}{c}{} & \multicolumn{1}{c}{} & $\tau = 200$ & -0.469\% & 0.06\% & 0.05\% & -0.01\% \\
\multicolumn{1}{c}{} &       &       &       &       &       &  \\
\multicolumn{1}{c}{} & \multicolumn{1}{c}{\multirow{2}[0]{*}{\textbf{RMSE}}} & $\theta $ &  \textless 1.0E-03 &  \textless 1.0E-03 &  \textless 1.0E-03 &  \textless 1.0E-03 \\
\multicolumn{1}{c}{} & \multicolumn{1}{c}{} & $\tau = $ & 6.2E+00 & 1.5E+00 & 6.1E-02 & 1.4E-01 \\
\multicolumn{1}{c}{} & \multicolumn{1}{c}{} &       &       &       &       &  \\
\multicolumn{1}{c}{} & \multicolumn{1}{c}{\multirow{2}[0]{*}{\textbf{Std dev}}} & $\theta $ & 2.7E-02 & 1.8E-02 & 1.2E-02 & 9.8E-03 \\
\multicolumn{1}{c}{} & \multicolumn{1}{c}{} & $\tau $ & 4.3E+01 & 3.0E+01 & 1.3E+01 & 9.2E+00 \\
\multicolumn{1}{c}{} &       &       &       &       &       &  \\
\multicolumn{1}{c}{} &       & $\overline{se(\theta)}$  & 1.1E-02 & 1.1E-02 & 5.2E-03 & 3.7E-03 \\
      &       & $\bar{\gamma}$  & 1.7E+08 & 1.7E+08 & 1.6E+08 & 1.6E+08 \\
      &       &       &       &       &       &  \\[2.5mm]
\hline
\end{tabular}%
 \label{tab:simulation_second_order}%
\end{table}%
The biases and the root mean squared errors of the estimates of the ODE parameters and of the state function at $t = 0$ are fairly small and tend to decrease with sample size. 
The estimation performance indicators for the constrained procedure are shown in the lower part of the table. 
It appears that the introduction of the initial conditions ensures, on average, more precise and robust estimates.
\section{Two real(istic) examples}
\label{section:application}
In this section the QL-ODE-P-spline procedure is illustrated by analyzing two real(isitc) extra examples. As first problem we consider a simulated experiment inspired by the heartbeat phenomenological model proposed by \cite{dossantosetal2004}. In the second example, (a subset of) the well known Canadian lynx-hare abundance data \citep{elton1942, holling1959} are analyzed by estimating the parameter and the solution of a Lotka-Volterra system of equations via QL-ODE-P-splines.
\subsection{A realistic model for the analysis of heartbeat signals}
\label{section:application_1}
\begin{figure}
\includegraphics[width = 1\textwidth]{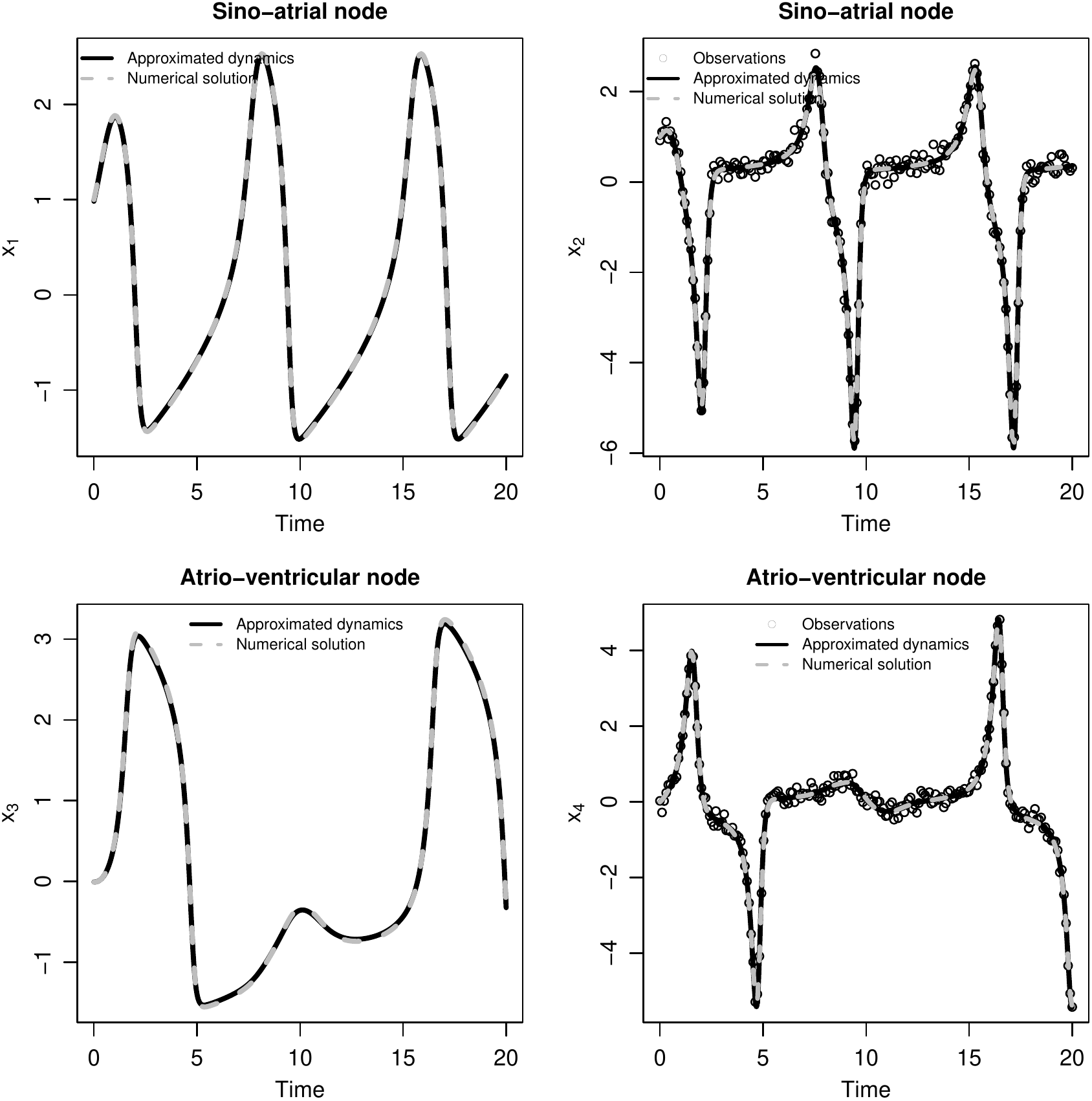}
\caption{The first row of the plot matrix depicts a simulated SA node (only the first derivative of the signal is observed) while the second shows the hypothetical voltage measured in the AV node (only the first derivative of the signal is observed). Two hundred measurements per node have been simulated.}
\label{fig:ECG_signal}
\end{figure}

The normal cardiac rhythm depends on the aggregate of cells in the right atrium defining the sino-atrial (SA) node which constitutes the normal pacemaker. The SA node generates electrical impulses that spread to the ventricules through the atrial musculature and conducting tissues, the AV node. The idea of treating the heart system using differential models dates back to the work by \cite{vanderpol1928}. In their pioneering paper the authors proposed to model the SA/AV system by a coupled set of electronic systems exhibiting relaxation oscillations. The \cite{vanderpol1928} model represents a convenient description of the hearth dynamics due to its parametric simplicity and ability to cover complex periodicity. Successively, other systems have been proposed generalizing this original framework in order to better synthesize the biological complexity of the phenomenon under consideration (see \cite{formaggia2010} for more details).\\ 
Among all the alternatives proposed in the specialized literature, we adopt the following system of coupled oscillators \citep{dossantosetal2004}: 
\[
\left\{
\begin{array}{ll}
\displaystyle \frac{d x_{1}}{d t}\left(t\right) = x_{2}\left(t\right)\\[0.3cm]
\displaystyle \frac{d x_{2}}{d t}\left(t\right) = \kappa (x_{1}\left(t\right) - w_{1}) (x_{1}\left(t\right) - w_{2}) x_{2}\left(t\right) - b_{1} x_{1}\left(t\right) + c_{1} (x_{3}\left(t\right) - x_{1}\left(t\right))\\[0.3cm] 
\displaystyle  \frac{d x_{3}}{d t}\left(t\right) = x_{4}\left(t\right)\\[0.3cm]
\displaystyle  \frac{d x_{4}}{d t}\left(t\right) = \kappa (x_{3}\left(t\right) - w_{1}) (x_{3}\left(t\right) - w_{2}) x_{4}\left(t\right) - b_{2} x_{3}\left(t\right) + c_{2} (x_{1}\left(t\right) - x_{3}\left(t\right))
\end{array}
\right.
\]
where $(x_{1}, x_{2})$ and $(x_{3}, x_{4})$ describe the electrical activity in the SA (sino-atrial) and AV (atrio-ventricular) nodes respectively. Parameters $b_{1}$ and $b_{2}$ determine the normal mode frequencies and $(w_{1}, w_{2})$ characterize the non-linearity of the system. A central role is played by parameters $(c_{1}, c_{2})$ determining the kind of diffusivity described by the coupled oscillators. For $c_{1} = c_{2} = 0$ the SA node undergoes periodic regular oscillations, while for non-null $c_{1} < c_{2}$ the system describes a bidirectional asymmetric coupling.\\
The right panels in Figure~\ref{fig:ECG_signal} depict the simulated measurements (dots) and the numerical solution of the system (dashed lines) together with the state function approximated via the QL-ODE-P-spline method (black full lines). The data have been obtained by perturbing $150$ uniformly sampled values of the first derivatives of the numerical solutions of the system with a zero mean Gaussian noise. The system has been solved numerically for $x_{1}(0) = 1, x_{2}(0) = 1, x_{3}(0) = \mbox{-1.0E-02}, x_{4}(0) = \mbox{-1.0E-02}$ taking $t \in (0, 20)$. Table~\ref{tab:ECG_signal} shows the estimated parameters and their standard errors. These results have been obtained using fourth order B-splines built on 250 equidistant internal knots. 
From Figure~\ref{fig:ECG_signal} and Table~\ref{tab:ECG_signal} one can appreciate the quality of the performances achieved by the QL-ODE-P-spline procedure.
\begin{table}
\centering
 \caption{ODE and precision parameter estimated for the observed coupled Van der Pol system \citep{dossantosetal2004}. In the first column the true values of the ODE and precision parameters are indicated.}
	\label{tab:ECG_signal}
    \begin{tabular*}{0.5\columnwidth}{@{\extracolsep{\fill} } r c c}
    \textbf{True parameters} & \textbf{Estimates} & \textbf{S. E.} \\
		\hline
    $b_{1} = 1.5$ & 1.509 & 0.016 \\
    $b_{2} = 0.1$ & 0.128 & 0.018 \\
    $\kappa = -1.8$ & -1.890 & 0.013 \\
    $w_{1} = -0.2$ & -0.201 & 0.017 \\
    $w_{2} = 2$ & 1.972 & 0.032 \\
    $c_{1} = 0.0$ & 0.004 & 0.010 \\
    $c_{2} = 0.55$ & 0.533 & 0.017 \\
		$\tau = 50$ & 49.73 &    \\
    $\widehat{\gamma}$ & 9.52E+03 &   \\
		\hline
    \end{tabular*}
\end{table}

\subsection{Canadian lynx and snowshoe hare life cycles}
\label{section:application_2}
\begin{figure}
\includegraphics[width = 1\textwidth]{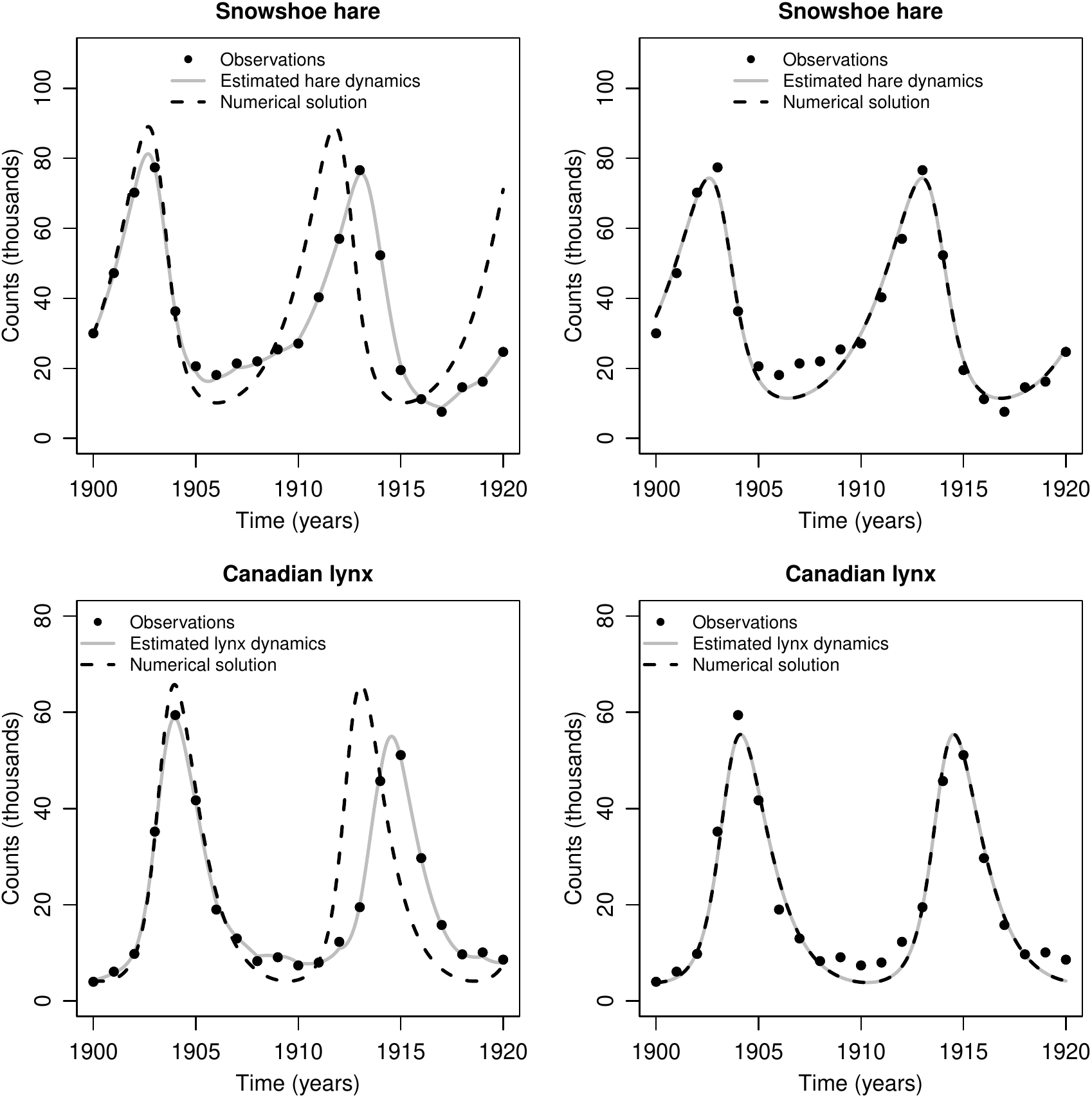}
\caption{Raw data and estimates obtained for the Canadian lynx vs snowshoe hare predator-pray dynamics. In this example we take into account the yearly observations recorded between 1900 and 1920. These estimates have been obtained taking $\gamma^{(0)} = 10^{-3}$ (first column of the figure) or $\gamma^{(0)} = 10^{3}$ (second column of the figure).}
\label{fig:lotka_volterra}
\end{figure}

In this section, we analyze the well known Canadian lynx and snowshoe hare data \citep{elton1942, holling1959}. The lynx and hare abundances have been collected from historical pelt-trading records of the Hudson Bay Company in a period between 1862 and 1930. Here, we consider a subset of the original sample analyzing the abundances observed between 1900 and 1920 (see figure~\ref{fig:lotka_volterra}). For these data the standard Lotka-Volterra system of ODEs can be estimated to obtain a meaningful description the life cycles of the two populations: 
\[
\left\{
\begin{array}{l l}
\displaystyle \frac{d x_{1}}{d t}\left(t\right) = x_{1}\left(t\right) [\beta - \zeta x_{2}\left(t\right)]\\[0.3cm]
\displaystyle \frac{d x_{2}}{d t}\left(t\right) = -x_{2}\left(t\right)[\delta - \eta x_{1}\left(t\right)]\\
\end{array}
\right.
\]
where $(x, y)$ indicate the prey (hare) and the predator (lynx) abundances respectively, $\beta$ and $\eta$ are the intrinsic growth rates of the prey and predator populations and $(\zeta, \delta)$ are the predator-pray meeting and the predator death rates respectively.\\
The parameter estimates obtained with the quasilinearized ODE-P-splines are given in Table~\ref{tab:result_lotka_volterra}. These results were computed using cubic B-splines defined on a set of 200 internal knots.
The initial values for the ODE and precision parameters, $\boldsymbol{\theta}^{(0)} = (0.547, 0.028, 0.843, 0.026)$ and $\tau^{\left(0\right)} = 0.01$, were obtained using nonlinear least squares with $x_{1}$  and $x_{2}$ approximated using standard P-splines \citep{eilersandmarx1996}.
Figure~\ref{fig:lotka_volterra} shows the approximated state functions comparing them with the raw measurements and the numerical solutions of the ODE system (black dashed lines). The numerical solutions have been computed plugging the ODE parameters and initial values estimated via QL-ODE-P-spline in the system and solving the ODEs through a Runge-Kutta scheme.\\ 
The unknown parameters and state functions have been estimated taking two initial values for the ODE-compliance parameter: one indicating (relatively) low initial compliance to the ODE model ($\gamma^{\left(0\right)} = 10^{-3}$) and one indicating (relatively) high initial compliance to the ODE system ($\gamma^{\left(0\right)} = 10^{3}$). In the first case the final smoother overfits the data by giving a low weight to the differential model (first column of Figure~\ref{fig:lotka_volterra}). Then, the (numerical) solutions obtained by plugging the estimated $\boldsymbol{\theta}$ in the system of ODEs are not in agreement with the observed data. On the contrary, starting with a large compliance parameter, a smoother consistent with the ODE model is estimated (second column of Figure~\ref{fig:lotka_volterra}): the approximated state functions (solid gray lines) and the numerical solutions (black dashed lines) overlap. In our opinion, these estimates guarantee a good description of the data even if the approximated state functions do not catch the trend observed in the years between 1905 and 1910. This miss-fitting effect could be a consequence of the poor flexibility of the Lotka-Volterra model. On the other hand we consider it a small price to be payed to preserve model parsimony and interpretability. \\
This fitting issue could also be related to the data collection process. Indeed, data before 1903 represent fur trading records while after 1903 they have been derived from questionnaires \citep{zhangetal2007}. This might have influenced the observations in the years between the two peaks. Finally, according to our experience, we advise to start the estimation process by taking a reasonably large $\gamma^{(0)}$ parameter if the research interest is focused on the estimation of the ODE parameters and the approximation of the associated state functions.

\begin{table}
\centering
  \caption{Estimated parameters of the Lotka-Volterra system defined for the lynx-hare data example. These results have been obtained taking $\gamma^{(0)} = 10^{-3}$ and $\gamma^{(0)} = 10^{3}$. The optimal ODE compliance parameters have been found equal to $10^{-1}$ and 2.71E+06 respectively.}
	\label{tab:result_lotka_volterra}
\begin{tabular*}{0.75\columnwidth}{@{\extracolsep{\fill} } r c c}
\textbf{Parameter}  &  \textbf{Estimates} ($\gamma^{(0)} = 10^{-3}$) &  \textbf{S. E.}  ($\gamma^{(0)} = 10^{-3}$)  \\
\hline
$\beta$ & 0.619 &  9.95E-02  \\
$\zeta$ & 0.028 &  3.66E-03   \\
$\delta$ & 0.971 &2.04E-01  \\
$\eta$ & 0.027 & 5.06E-03 \\[0.15cm]
$\widehat{\tau}$ & 0.635 &      \\
  &  &     \\
 \textbf{Parameter} &  \textbf{Estimates} ($\gamma^{(0)} = 10^{3}$) &  \textbf{S. E.}($\gamma^{(0)} = 10^{3}$) \\
\hline
$\beta$ & 0.481 & 3.70E-02      \\
$\zeta$ & 0.025 & 2.00E-03 \\
$\delta$ & 0.927 & 7.60E-02  \\
$\eta$ &0.028 & 2.00E-03    \\[0.15cm]
$\widehat{\tau}$ & 3.863 &\\
\hline
\end{tabular*}
\end{table}

\section{Discussion}
\label{section:discussion}
In this paper we have introduced the quasilinearized (QL-) ODE-P-spline approach as a tool for the estimation of the parameters defining (systems of) nonlinear ODEs and for the approximation of their state functions starting from a set of (noisy) measurements. This methodology is inspired by the generalized profiling framework of \cite{ramsayetal2007}. The estimation procedure exploits a penalized likelihood formulation to build a compromise between a B-spline based description of the observations and the B-spline collocation solution of the ODE. The estimation process alternates two steps: 1) estimation of the optimal spline coefficients given the other unknowns; 2) estimation of the ODE parameters treating the spline coefficients as nuisance parameters. The compromise between data smoothing and compliance to the differential model is tuned by an ODE-compliance parameter to be selected in an upper optimization level. Dealing with nonlinear ODEs, the application of the original proposal of \cite{ramsayetal2007} is involved requiring a nonlinear least squares step and the use of the implicit function theorem both for point and interval estimates. This is a consequence of the implicit relationship between the spline coefficients and the other unknowns. \\
The introduction of a quasilinearization step overcomes these hitches leading to a valuable simplification of the estimation procedure. The quasilinearization of the ODE-based penalty term \citep{bellmanandkalaba1965} permits to analytically link the spline coefficients to the differential parameters and makes possible the selection of the ODE-compliance parameters using standard approaches. Indeed, within our quasilinearized spline based framework, the estimation process reduces to a conditionally linear problem for the optimization of the spline coefficients. We suggest to view the ODE-compliance parameter as the ratio of the noise and of the penalty variances and to estimate it using the EM-type approach proposed by \citet{schall1991}. The quasilinearized framework also facilitates the introduction of state (initial or boundary value) conditions. As discussed in Section~\ref{subsection:differential_conditions}, these constraints can be imposed using either extra $L_{2}$ penalties or Lagrange multipliers. \\
According to Algorithm~\ref{algo_ql_ode_p_spline}, the quasilinearized estimation process requires initial values $\boldsymbol{\alpha}^{\left(0 \right)}$ for the spline coefficients. Good initial spline coefficients can be obtained smoothing the raw data with standard P-splines \citep{eilersandmarx1996} even if, according to our experience, the choice of $\boldsymbol{\alpha}^{(0)}$ hardly influences the quality of the final estimates (see Section~\ref{section:quasilinearized_ode_p_spline_approach}).\\
The performances of the proposed method have been evaluated through intensive simulations (Section~\ref{section:simulation}). Two simulation studies have been conducted by generating data from a simple first order ODE in the first case and by considering a relaxation oscillator governed by the Van der Pol equation in the second case. The first ODE model, having an explicit solution, made possible a precise comparison between the estimates obtained via QL-ODE-P-splines and by a nonlinear least squares inversion of the analytic solution \citep{hotelling1927, biegleretal1986}. The simulation results confirm that our approach ensures performances comparable with those obtained via NLS for known or unknown state conditions. The high quality of the QL-ODE-P-spline estimates can also be appreciated by looking at the simulation results obtained in the framework of the more complex Van der Pol differential problem. \\
In Section~\ref{section:application}, the quasilinearized spline based methodology has been applied in two real(istic) data analyzes. In the first example we have analyzed a set of 150 synthetic SA/AV voltage measurements generated perturbing with a Gaussian random noise the first derivatives of the (numerical) solutions of an unforced system of coupled Van der Pol equations \citep{dossantosetal2004}. In the second example we have analyzed the (yearly) lynx-hare abundances \citep{maclulich1937, elton1942} recorded in Canada between 1900 and 1920. In the latter case, we used a QL-ODE-P-spline approach with a penalty based on a Lotka-Volterra system. The estimation has been conducted by taking either a small or a large initial value for the ODE-compliance parameter. Starting with a large initial compliance parameter ($\gamma^{(0)} = 10^{3}$) ensured final estimates for the state functions and ODE parameters consistent with the hypothesized model, while a low initial compliance parameter ($\gamma^{(0)} = 10^{-3}$) led to overfitting state functions corresponding to inappropriate estimates for the ODE parameters. This is probably due to the inability of the ODE model to describe the measurement trend in the period 1905-1910. A small value for the ODE compliance parameter provides a better fit but makes the interpretation of the final ODE parameter estimates awkward.\\
Our future research will focus on possible extensions of the QL-ODE-P-spline framework. First of all, we think that the linearized ODE penalty and the profiling estimation settings can be adapted to hierarchical settings with random ODE parameters. Second, the presented approach can be also adapted to analyze (noisy) realizations of dynamic systems evolving in space and time and described by nonlinear partial differential equations (PDEs). Finally, the introduction of a quasilinearization step should facilitate the generalization of the approach proposed by \cite{jaegerandlambert2013} to analyze nonlinear differential systems in a Bayesian framework.
\section*{Acknowledgements}
The authors acknowledge financial support from IAP research network P7/06 
of the Belgian Government (Belgian Science Policy), and from the contract 
`Projet d'Actions de Recherche Concert\'ees' (ARC) 11/16-039 of the 
`Communaut\'e fran\c{c}aise de Belgique', granted by the `Acad\'emie universitaire Louvain'.
\newpage
\appendix
\setcounter{equation}{0}
\renewcommand{\theequation}{\thesection.\arabic{equation}}
\section{Construction of the quasilinearized penalty matrix in a general system of nonlinear ODEs}
\label{section:app_A}
In this appendix we define the matrix $\boldsymbol{R}\left(\boldsymbol{\theta}, \boldsymbol{\gamma}, \boldsymbol{\alpha}^{(i)}\right)$, the vector $\boldsymbol{r}\left(\boldsymbol{\theta}, \boldsymbol{\gamma}, \boldsymbol{\alpha}^{(i)}\right)$ and the constant $\boldsymbol{l}\left(\boldsymbol{\theta}, \boldsymbol{\gamma}, \boldsymbol{\alpha}^{(i)}\right)$ involved in the linearized ODE-penalty. Consider a general system of ODEs:
\begin{equation}\label{eq:general_system}
\left\{
\begin{array}{ll}
\displaystyle \frac{d x_{1}}{d t}\left(t\right) - f_{1}(t , \boldsymbol{x}, \boldsymbol{\theta}) = 0\\
\ \		\vdots\\
\displaystyle \frac{d x_{d}}{d t}\left(t\right) - f_{d}(t , \boldsymbol{x}, \boldsymbol{\theta}) = 0
\end{array}
\right.
\end{equation}
where $f_{j}(t , \boldsymbol{x}, \boldsymbol{\theta})$ is a known function of the state variables and of the unknown ODE parameters $\boldsymbol{\theta}$ that we consider fixed over time. With quasilinearization, given the estimate $\widetilde{\boldsymbol{x}}^{\left(i\right)}$ at step $i$, Eq.~(\ref{eq:general_system}) suggests to require that  $\widetilde{\boldsymbol{x}}^{\left(i + 1\right)}$ checks:
\[
\left\{
\begin{array}{ll}
\displaystyle \frac{d \widetilde{x}_{1}^{\left(i + 1\right)}}{d t}\left(t \right) - \sum_{k = 1}^{d} \displaystyle  \widetilde{x}_{k}^{\left(i + 1 \right)} \frac{\partial f_{1}}{\partial x_{k}} \left(t, \widetilde{\boldsymbol{x}}^{(i)}, \boldsymbol{\theta}\right) - f_{1}\left(t, \widetilde{\boldsymbol{x}}^{(i)}, \boldsymbol{\theta}\right) + \sum_{k = 1}^{d} \displaystyle \widetilde{x}_{k}^{\left(i  \right)} \frac{\partial  f_{1}}{\partial x_{k}} \left(t, \widetilde{\boldsymbol{x}}^{(i)}, \boldsymbol{\theta}\right) = 0\\
\ \ \ \		\vdots\\
\displaystyle \frac{d \widetilde{x}_{d}^{\left(i + 1\right)}}{d t}\left(t \right) - \sum_{k = 1}^{d} \displaystyle  \widetilde{x}_{k}^{\left(i + 1 \right)} \frac{\partial f_{d}}{\partial x_{k}} \left(t, \widetilde{\boldsymbol{x}}^{(i)}, \boldsymbol{\theta}\right) - f_{d}\left(t, \widetilde{\boldsymbol{x}}^{(i)}, \boldsymbol{\theta}\right) + \sum_{k = 1}^{d} \displaystyle  \widetilde{x}_{k}^{\left(i  \right)} \frac{\partial f_{d}}{\partial x_{k}} \left(t, \widetilde{\boldsymbol{x}}^{(i)}, \boldsymbol{\theta}\right) = 0
\end{array}
\right.
\]
If $\boldsymbol{\alpha}_{j}$ denotes the vector of spline coefficients to be estimated in the $j$th state function, the updated $j$th penalty term at iteration $i+1$ is given by:
\begin{eqnarray*}
{PEN}_{j}\left(\boldsymbol{\alpha}_{j}^{\left(i+1\right)}|\boldsymbol{\alpha}_{j}^{\left(i\right)}\right) & \approx & \int \left(\frac{d \widetilde{x}_{j}^{\left(i+1\right)}}{dt}\left(t\right) - f_{j}\left(t, \widetilde{\boldsymbol{x}}^{\left(i\right)}, \boldsymbol{\theta}\right) \right. \\ 
&-& \left. \sum_{k = 1}^{d}{\left(\texttt{}\widetilde{x}_{k}^{\left(i+1\right)} - \widetilde{x}_{k}^{\left(i\right)}\right) \displaystyle\frac{\partial f_{j}}{\partial x_{k}}\left(t, \widetilde{\boldsymbol{x}}^{\left(i\right)}, \boldsymbol{\theta}\right) }\right)^{2}dt , \\ 
&\approx &  \int \left( \left[\boldsymbol{\alpha}_{j}^{(i + 1)}\right]^{\top} \boldsymbol{b}^{\left(1\right)}_{j}\left(t\right) \right. \\
 & -& \left. \sum_{k = 1}^{d} \left[\boldsymbol{\alpha}^{\left(i + 1\right)}_{k}\right]^{\top} \boldsymbol{b}^{\left(0\right)}_{j} \left( t \right)\frac{\partial f_{i}}{\partial x_{k}} \left(t, \widetilde{\boldsymbol{x}}^{\left(i\right)}, \boldsymbol{\theta} \right) - v_{j}\left(t\right) \right)^{2} dt
\end{eqnarray*}
where the integration is performed over $\left[0, T\right]$, $v_{j}\left(t \right) = f_{j} \left(t, \widetilde{\boldsymbol{x}}^{\left(i\right)}, \boldsymbol{\theta} \right) - \displaystyle \sum_{k = 1}^{d} \widetilde{x}^{\left(i \right)}_{k}\left(t \right) \frac{\partial f_{j}}{\partial x_{k}} \left(t, \widetilde{\boldsymbol{x}}^{\left(i\right)}, \boldsymbol{\theta} \right) $ and $\boldsymbol{b}^{\left(0\right)}_{j}(t)$, $\boldsymbol{b}^{\left(1\right)}_{j}(t)$ are, respectively, $K_{j}$-dimensional vectors of B-spline functions and their derivative such that $\widetilde{x}^{(i)}_{j}(t) = \left[\boldsymbol{b}^{\left(0 \right)}_{j}(t) \right]^{\top} \boldsymbol{\alpha}^{(i)}_{j} $and  $\displaystyle \frac{d \widetilde{x}^{(i)}_{j}}{d t}(t) = \left[\boldsymbol{b}^{\left(1 \right)}_{j}(t) \right]^{\top} \boldsymbol{\alpha}^{(i)}_{j} $. 
The $j$th element of the penalty can be computed as:
\begin{equation}\label{eq:jth_penalty}
\begin{array}{l l l}
\displaystyle PEN_{j} &=& \left[\boldsymbol{\alpha}_{j}^{\left(i + 1\right)} \right]^{\top} \left( \displaystyle \int \boldsymbol{b}^{\left(1\right)}_{j}\left(t\right) \left[\boldsymbol{b}^{\left(1\right)}_{j}\left(t\right) \right]^{\top} dt \right)\boldsymbol{\alpha}_{j}^{\left(i + 1\right)} \\
 \displaystyle &-& \left[\boldsymbol{\alpha}_{j}^{\left(i + 1\right)} \right]^{\top}\left(\displaystyle  \displaystyle \int \frac{\partial f_{j}}{\partial x_{1}} \left(t, \widetilde{\boldsymbol{x}}^{\left(i\right)}, \boldsymbol{\theta}\right) \boldsymbol{b}^{\left(1\right)}_{j}\left(t\right) \left[\boldsymbol{b}^{\left(0\right)}_{1}\left(t\right)\right]^{\top} dt, \cdots , \right.\\
\displaystyle && \left. \displaystyle \int \frac{\partial f_{j}}{\partial x_{d}} \left(t, \widetilde{\boldsymbol{x}}^{\left(i\right)}, \boldsymbol{\theta}\right)  \boldsymbol{b}^{\left(1\right)}_{j}\left(t\right) \left[\boldsymbol{b}^{\left(0\right)}_{d}\left(t\right)\right]^{\top}dt \right) \boldsymbol{\alpha}^{\left(i + 1\right)}\\
\displaystyle & - & \left[\boldsymbol{\alpha}^{\left(i + 1\right)} \right]^{\top}
\left( 
\begin{array}{c}
\displaystyle \int \frac{\partial f_{j}}{\partial x_{1}} \left(t, \widetilde{\boldsymbol{x}}^{\left(i\right)}, \boldsymbol{\theta}\right) \boldsymbol{b}^{\left(0\right)}_{1}\left(t\right) \left[\boldsymbol{b}^{\left(1\right)}_{j}\left(t\right)\right]^{\top} dt\\
 \vdots\\
\displaystyle \int \frac{\partial f_{j}}{\partial x_{d}} \left(t, \widetilde{\boldsymbol{x}}^{\left(i\right)}, \boldsymbol{\theta}\right) \boldsymbol{b}^{\left(0\right)}_{d}\left(t\right) \left[\boldsymbol{b}^{\left(1\right)}_{j}\left(t\right)\right]^{\top} dt
\end{array}
\right) \boldsymbol{\alpha}_{j}^{\left(i + 1 \right)}\\
\displaystyle &+&  \left[\boldsymbol{\alpha}^{\left(i + 1\right)} \right]^{\top}
\left(
\begin{array}{c c c c}
\displaystyle c_{j_{11}} & c_{j_{12}} & \cdots&  c_{j_{1d}}\\
\displaystyle c_{j_{21}} & \dots& \dots & c_{j_{2d}}\\
\displaystyle \vdots & \vdots & \ddots & \vdots  \\
\displaystyle c_{j_{d1}}& \dots & \dots & c_{j_{dd}}
\end{array}
\displaystyle \right) \boldsymbol{\alpha}^{\left(i + 1\right)}\\
\displaystyle &+& 2 \left[\boldsymbol{\alpha}^{\left(i + 1\right)} \right]^{\top} 
\left(
\begin{array}{c}
\displaystyle \int{ v_{j}\left(t\right) \frac{\partial f_{j}}{\partial x_{1}} \left(t, \widetilde{\boldsymbol{x}}^{\left(i\right)}, \boldsymbol{\theta} \right) \boldsymbol{b}^{\left( 0 \right)}_{j} \left(t\right)}dt\\
 \vdots\\
\displaystyle \int{ v_{j}\left(t\right) \frac{\partial f_{j}}{\partial x_{d}} \left(t, \widetilde{\boldsymbol{x}}^{\left(i\right)}, \boldsymbol{\theta} \right) \boldsymbol{b}^{\left( 0 \right)}_{j} \left(t\right)}dt
\end{array}
\right)\\
&-& 2  \displaystyle \left( \int{ v_{j}\left(t\right)  \boldsymbol{b}^{\left( 1 \right)}_{j} \left(t\right)} dt \right)^{\top} \boldsymbol{\alpha}^{\left(i + 1\right)} +  \int{ v^{2}_{j}\left(t\right) }dt,
\end{array}
\end{equation}
with
\[
c_{j_{kl}} = \int \frac{\partial f_{j}}{\partial x_{k}} \left(t, \widetilde{\boldsymbol{x}}^{\left(i\right)}, \boldsymbol{\theta} \right) \frac{\partial f_{j}}{\partial x_{l}} \left(t, \widetilde{\boldsymbol{x}}^{\left(i\right)}, \boldsymbol{\theta} \right) \boldsymbol{b}^{\left(0\right)}_{k}\left(t\right) \left[\boldsymbol{b}^{\left(0\right)}_{l}\left(t\right)\right]^{\top} dt.
\]
The overall penalty term $PEN^{\left(i + 1\right)} = \sum_{j = 1}^{d} \gamma_{j} PEN^{\left(i + 1\right)}_{j}$, at step $i + 1$ is therefore equal to:
\[
PEN^{\left(i + 1 \right)} = \left[\boldsymbol{\alpha}^{\left(i + 1\right)}\right]^{\top} \boldsymbol{R}\left(\boldsymbol{\theta}, \boldsymbol{\gamma}, \boldsymbol{\alpha}^{(i)}\right) \boldsymbol{\alpha}^{\left(i + 1\right)} + 2 \left[\boldsymbol{\alpha}^{\left(i + 1\right)}\right]^{\top} \boldsymbol{r}\left(\boldsymbol{\theta}, \boldsymbol{\gamma}, \boldsymbol{\alpha}^{(i)}\right) + l \left(\boldsymbol{\theta}, \boldsymbol{\gamma}, \boldsymbol{\alpha}^{(i)}\right).
\]
The matrix $\boldsymbol{R}\left(\boldsymbol{\theta}, \boldsymbol{\gamma}, \boldsymbol{\alpha}^{(i)}\right) $ has dimension $K \times K$, is symmetric and block-diagonal. The block $(k, l)$ has dimension $K_{k} \times K_{l}$
and is computed as:
\begin{eqnarray*}
&& \delta_{kl} \gamma_{k} \int \boldsymbol{b}^{\left(1\right)}_{k}\left(t\right) \left[\boldsymbol{b}^{\left(1\right)}_{k}\left(t\right) \right]^{\top} dt - \gamma_{k} \int \frac{\partial f_{j}}{\partial x_{k}} \left(t, \widetilde{\boldsymbol{x}}^{\left(i\right)}, \boldsymbol{\theta}\right) \boldsymbol{b}^{\left(1\right)}_{k}\left(t\right) \left[\boldsymbol{b}^{\left(0\right)}_{l}\left(t\right)\right]^{\top} dt \\
& & - \gamma_{l} \int \frac{\partial f_{j}}{\partial x_{l}} \left(t, \widetilde{\boldsymbol{x}}^{\left(i\right)}, \boldsymbol{\theta}\right) \boldsymbol{b}^{\left(0\right)}_{l}\left(t\right) \left[\boldsymbol{b}^{\left(1\right)}_{k}\left(t\right)\right]^{\top} dt + \sum_{i = 1}^{d} \gamma_{i} c_{i_{kl}},
\end{eqnarray*}
where $k,\ l \in \{1 ,\dots ,d \}$ and $\delta_{kl} = 1$ if $k = l$ and zero otherwise. The vector $\boldsymbol{r}\left(\boldsymbol{\theta}, \boldsymbol{\gamma}, \boldsymbol{\alpha}^{(i)}\right)$ has length $K$ with $k$th component given by $\displaystyle \sum_{j = 1}^{d} \gamma_{j} \left( \int{ v_{j}\left(t\right) \frac{\partial f_{j}}{\partial x_{1}} \left(t, \widetilde{\boldsymbol{x}}^{\left(i\right)}, \boldsymbol{\theta} \right) \boldsymbol{b}^{\left( 0 \right)}_{k} \left(t\right)}dt \right) - \gamma_{k} \int{ v_{k}\left(t\right)  \boldsymbol{b}^{\left( 1 \right)}_{k} \left(t\right)} dt $. Finally the constant $l \left(\boldsymbol{\theta}, \boldsymbol{\gamma}, \boldsymbol{\alpha}^{(i)}\right) $ in the penalty is equal to $\displaystyle \sum_{j = 1}^{d} \gamma_{j}  \int{ v^{2}_{j}\left(t\right) }dt $.

\end{document}